\documentclass[12pt,preprint]{aastex}
\usepackage{amsfonts}
\usepackage{apjfonts}
\usepackage{amssymb, natbib, amsmath, epsfig, epsf, lscape, rotating}
\newcommand{\flux}{erg~s$^{-1}$~cm$^{-2}$}
\newcommand{\lum}{erg~s\ensuremath{^{-1}}}
\newcommand{\lbol}{\ensuremath{L\mathrm{_{bol}}}}

\newcommand{\ledd}{\ensuremath{L\mathrm{_{Edd}}}}
\newcommand{\lratio}{\ensuremath{\lbol/\ledd}}

\newcommand{\msun}{\ensuremath{M_{\odot}}}
\newcommand{\kms}{\ensuremath{\mathrm{km~s^{-1}}}}
\newcommand{\mbh}{\ensuremath{M_\mathrm{BH}}}

\newcommand{\rs}{\ensuremath{r_{\rm \scriptscriptstyle S}}}
\newcommand{\pnull}{\ensuremath{P_{\mathrm{null}}}}

\newcommand{\chisq}{\ensuremath{\chi^2}}

\newcommand{\ha}{H\ensuremath{\alpha}}
\newcommand{\hb}{H\ensuremath{\beta}}

\newcommand{\bha}{H\ensuremath{\alpha ^{\rm B}}}

\newcommand{\hii}{H\,{\footnotesize II}}
\newcommand{\nii}{[N\,{\footnotesize II}]}
\newcommand{\sii}{[S\,{\footnotesize II}]}
\newcommand{\oi}{[O\,{\footnotesize I}]}
\newcommand{\oiii}{[O\,{\footnotesize III}]}

\newcommand{\feii}{{\rm Fe\,{\footnotesize II}}}

\newcommand{\mgii}{Mg\,{\footnotesize II}}

\newcommand{\civ}{C\,{\footnotesize IV}}

\newcommand{\caii}{Ca\,{\footnotesize II}}

\def\lax{{$\mathrel{\hbox{\rlap{\hbox{\lower4pt\hbox{$\sim$}}}\hbox{$<$}}}$}}
\def\gax{{$\mathrel{\hbox{\rlap{\hbox{\lower4pt\hbox{$\sim$}}}\hbox{$>$}}}$}}
\slugcomment{To appear in {\it The Astrophysical Journal}}
\shorttitle{Seyfert 1 Nuclei with Low-mass Black Holes} %
\shortauthors{X.-B.~Dong et~al.}

\begin{document}

\title{A Uniformly Selected Sample of Low-mass Black Holes in Seyfert\,1 Galaxies}

\author{%{\color{green} \it(tentatively)}
Xiao-Bo~Dong\altaffilmark{1,2}, Luis~C.~Ho\altaffilmark{2},
Weimin~Yuan\altaffilmark{3}, Ting-Gui~Wang\altaffilmark{1},
Xiaohui~Fan\altaffilmark{4}, Hongyan~Zhou\altaffilmark{1,5}, and
Ning~Jiang\altaffilmark{1} }

\altaffiltext{1}{Key laboratory for Research in Galaxies and
Cosmology, Department of Astronomy, The University of Sciences and
Technology of China, Chinese Academy of Sciences, Hefei, Anhui
230026, China; \mbox{xbdong@ustc.edu.cn} } %
\altaffiltext{2}{The Observatories of the Carnegie Institution for
Science, 813 Santa Barbara Street, Pasadena, CA 91101, USA;
\mbox{lho@obs.carnegiescience.edu} } %

\altaffiltext{3}{National Astronomical Observatories, Chinese
Academy of Sciences, Beijing, 100012, China; \mbox{wmy@nao.cas.cn} } %
\altaffiltext{4}{Steward Observatory, The University of Arizona,
Tucson, AZ 85721, USA; \mbox{fan@as.arizona.edu} } %
\altaffiltext{5}{Polar Research Institute of China, Jinqiao Rd. 451,
Shanghai 200136, China; \mbox{zhouhongyan@pric.gov.cn}}

%Journal---
%2011 april 28, xbdong
%2011 may 16, xbdong
%\email{\scriptsize \it 10 June, 2011}
%\email{\scriptsize \it \today}
\email{\scriptsize \it Received 2011 October 10}  %; accepted 2011 May 3}

\begin{abstract}
We have conducted a systematic search of low-mass black holes (BHs)
in active galactic nuclei (AGNs) with broad \ha\ emission lines,
aiming at building a homogeneous sample that is more complete than
previous ones for fainter, less highly accreting sources.  For this
purpose, we developed a set of elaborate, automated selection
procedures and applied it uniformly to the Fourth Data Release of
the Sloan Digital Sky Survey.  Special attention is given to
AGN--galaxy spectral decomposition and emission-line deblending. We
define a sample of 309 type 1 AGNs with BH masses in the range $8
\times 10^4$--$2 \times 10^6$~\msun\ (with a median of $1.2 \times
10^6$~\msun), using the virial mass estimator based on the broad
H$\alpha$ line.  About half of our sample of low-mass BHs differs
from that of Greene \& Ho, with 61 of them discovered here for the
first time. Our new sample picks up more AGNs with low accretion
rates: the Eddington ratios of the present sample range from
$\lesssim0.01$ to $\sim 1$, with 30\% below $0.1$. This suggests
that a significant fraction of low-mass BHs in the local Universe
are accreting at low rates. The host galaxies of the low-mass BHs
have luminosities similar to those of $L^*$ field galaxies, optical
colors of Sbc spirals, and stellar spectral features consistent with
a continuous star formation history with a mean stellar age of less
than 1~Gyr.
\end{abstract}


\setcounter{footnote}{0}
\setcounter{section}{0}

\section{Introduction}

Supermassive black holes (BHs) with masses $\mbh \gtrsim 10^6$ \msun\ reside
in the centers of galaxies with a spheroidal stellar component. Although their
formation mechanisms are still not well understood, correlations of their
masses with the host galaxy properties have been fairly well established (Magorrian
et al.  1998; Gebhardt et al. 2000; Ferrarese \& Merritt 2000).  However, far
less is known about lower mass BHs in the regime \mbh\ $\approx 10^3-10^6$
\msun, the so-called intermediate-mass BHs, presumably hosted by smaller
stellar systems such as dense star clusters and late-type or dwarf galaxies
(see, e.g., van~der~Marel 2004 for a review).  Intermediate-mass BHs are
important for studies of BH formation and growth, galaxy formation and
evolution, and active galactic nuclei (AGNs). In current models of galaxy
evolution in a hierarchical cosmology, supermassive BHs must have been built
up from accretion onto much smaller ``seeds,'' in conjunction with merging
with other BHs. These models also predict that smaller-scale structures form
at later times (cosmic ``downsizing''), and one might expect that seed
BHs in smaller stellar systems may not have had enough time to be fully grown.
A population of intermediate-mass BHs likely exists in the local
Universe. The mass function of the present-day intermediate-mass BHs and
their host galaxy properties can be used to discriminate between different
models for seed BHs and help shed light on the coevolution of BHs and
galaxies (e.g., Volonteri et~al. 2008).

Searching for intermediate-mass BHs by direct, dynamical measurements turns
out to be a difficult task. The gravitational sphere of influence
of these objects cannot be resolved using current facilities, except for a
handful of the nearest ($\lesssim 1$ Mpc) star clusters and galaxies.
A practicable approach is to search, through optical spectroscopic surveys, for
those BHs that are actively accreting and thus revealing themselves as AGNs,
for which their virial masses can be estimated using the empirical relations
using emission-line width and luminosity (e.g., Kaspi et~al. 2000; Greene \&
Ho 2005b). A prototype of such a kind of AGN is the Seyfert nucleus residing in
the nearby dwarf spiral (Sdm) galaxy NGC\,4395 (Filippenko \& Ho 2003), which
has a BH mass $3.6 \pm 1.1 \times 10^5$~\msun\ as measured via reverberation
mapping observations of the \civ\ $\lambda1549$ line (Peterson et~al. 2005).
Using spectroscopic data from the Sloan Digital Sky Survey (SDSS; York
et~al. 2000) Greene \& Ho (2004, 2007b) have conducted systematic
searches for AGNs with intermediate-mass BHs, leading to the
discovery of some 200 candidates. One noticeable feature of their
sample is the overall relatively high Eddington ratios (\lratio)
and a small number of objects with $\lratio <0.1$, a likely
consequence of their strict criteria for detecting the broad \ha\ line.

According to Greene \& Ho (2007a), the space density of massive BHs in AGNs
decreases toward the low-mass end ($\mbh \lesssim 2 \times 10^6$~\msun).%
\footnote{Following Greene \& Ho (2007b), hereinafter we refer to BHs with
$\mbh < 2 \times 10^6$~\msun\ at the centers of galaxies as ``low-mass'' or
``intermediate-mass'' BHs. We use the two terms interchangeably. Accordingly,
for the ease of narration, hereinafter we call AGNs hosting low-mass
BHs as low-mass AGNs, wherever it is not ambiguous.}
This may imply that intermediate-mass BHs are truly rare, or their
AGN phase has a short duty cycle. The former may be due either to
the decreasing bulge fraction in small galaxies and/or to an existing
lower limit on the possible mass of nuclear BHs (see, e.g., Haehnelt
et~al. 1998). From the standpoint of observations, however, it is
plausible that the apparent paucity may just be a selection effect. This is
because low-mass AGNs radiate at low luminosities, even if they are shining at
their maximum rate, and, even worse, detection of their continuum and line
emission is often susceptible to dilution by starlight of the host
galaxy due to the limited spatial resolution of observations. This may
explain the dearth of low-mass AGNs with $\lratio <0.1$ in the
sample of Greene \& Ho (2007a), who adopted very strict selection criteria
for detecting broad emission lines.  After all, low-mass AGNs with low
accretion rates do exist.  Depending on the adopted BH mass, NGC\,4395 has
\lratio\ $\approx 0.002-0.02$ (Filippenko \& Ho 2003; Peterson et~al. 2005),
and low Eddington ratios have been reported for smaller samples of low-mass
BHs in late-type galaxies based on infrared (Satyapal et~al. 2008) and X-ray
(Desroches \& Ho 2009) observations.

%% below: why still to compile a large sample in the optical
% xbdong, May 15, 2010

The above considerations suggest that it is desirable to compile a more
complete, homogeneous sample of low-mass AGNs, with emphasis on enlarging the
number of systems with low accretion. This would result in a less biased
distribution of accretion rates (as represented by \lratio), which is
important for a more complete picture of BH demographics (e.g., Schulze \&
Wisotzki 2010).  Moreover, low-\lratio, low-mass BHs are interesting in and of
themselves for AGN research, as they extend the dynamic range of physical
variables to extreme values.

% The above considerations suggests that it is desirable to compile a
% more complete, homogeneous sample of low-mass AGNs, with emphasis on
% enlarging the number of systems with low accretion. This would
% result in a less biased distribution of accretion rates (as
% represented by \lratio), which is important for a more complete
% picture of BH demographics (e.g., Schulze \& Wisotzki 2010).
% Moreover, low-accretion, low-mass AGNs similar to NGC~4395 are of
% special interest of their own. They are irreplaceable in studying
% the nuclear region of the host galaxies, as the AGN light is
% substantially reduced, which may shed light on the relationship
% between central BHs and host galaxies (N.~Jiang \etal\ 2011, in
% preparation; cf. Nowak et~al. 2010 for a detailed study on the
% nuclear substructures in nearby inactive galaxies). Besides, they
% are interesting in and of themselves for AGN research, as they
% extend the dynamic range of physical variables to extreme values.

%% xbdong, May 15, 2010
Optical spectroscopy is a powerful means of estimating BH masses for AGNs on a
large scale. However, like the proverbial finding a needle in a haystack,
searching for low-mass AGNs with low \lratio\ in the SDSS database is rather
challenging, since their spectra within the 3\arcsec-diameter fiber aperture
are generally dominated by starlight from the host galaxies.  To recover the
weak broad \ha\ line, the strongest permitted line in the optical region on
which we base our BH mass estimation, requires two essential steps---accurate
decomposition of AGN-starlight spectra and careful deblending of the broad \ha\
component from its associated narrow lines (narrow \ha\ and the
\nii\,$\lambda\lambda6548,6583$ doublet). Equally important is to design
carefully both a set of broad-line criteria and an automated selection
procedure to mine effectively the huge volume data set.

In this paper we present a sample of 309 Seyfert 1 galaxies hosting
a central BH of $\mbh < 2 \times 10^6$~\msun, uniformly selected
from the fourth data release (DR4; Adelman-McCarthy et~al. 2006) of
the SDSS.  We have carefully designed selection criteria and
spectral fitting methods aimed at detecting low-mass BHs with relatively
weak emission lines. Their Eddington ratios are found to range from
$\lesssim0.01$ to $\sim 1$, with 30\% of the objects having $\lratio
< 0.1$. The spectral fitting method is described in Section 2,
and the selection criteria and sample construction in Section 3.
Results of the spectral and imaging analyses are presented in
Section 4, followed by a summary in Section 5.
We use a cosmology with $H_0 = 70$~\kms~Mpc$^{-1}$, $\Omega_m
= 0.3$, and $\Omega_{\Lambda} = 0.7$.

%% Section 2
\section{Data and Processing Methods}

This section describes several technical aspects of our procedure for mining
the SDSS spectroscopic data to construct the
low-mass AGN sample, mainly concerning continuum and emission-line
fitting and assessment of the uncertainties of derived parameters.
As our data reduction and analysis methods are largely shaped by the
characteristics of the SDSS data, we
first give a brief description of the SDSS data (see
Stoughton et~al. 2002 for details).

\subsection{SDSS Data}
The SDSS is an imaging and spectroscopic survey, using a dedicated
2.5-m telescope to image one-quarter of the sky and to perform
follow-up spectroscopic observations. The imaging data were
collected in a drift-scan mode in five bandpasses ($u$, $g$, $r$,
$i$, and $z$) on nights of pristine conditions. The photometric
calibration is accurate to 5\%, 3\%, 3\%, 3\%, and 5\%,
respectively. Targets for spectroscopy were selected based on the
photometric data. Galaxy candidates were targeted as resolved
sources with $r$-band Petrosian magnitudes $< 17.77$; luminous red
galaxy candidates were specifically targeted by two different sets
of criteria with $r$-band Petrosian magnitude $< 19.2$ and 19.5,
respectively; low-redshift quasar candidates were selected according
to a set of color criteria with $i$-band PSF magnitude $< 19.1$;
high-redshift quasar candidates were selected according to a
different set of color criteria with $i$-band PSF magnitude $<20.2$.
Also targeted were unresolved objects with radio counterparts
detected by the FIRST survey (Becker et~al. 1995),
partial objects with X-ray counterparts detected by {\it ROSAT}\ (Voges et~al.
1999) and other serendipitous sources when free fibers
were available on a plate. Fibers that feed the SDSS spectrographs
subtend a diameter of 3\arcsec\ on the sky, which corresponds to 6.5
kpc at $z=0.1$. The nominal total exposure time for each plate is 45
minutes, which typically yields a signal-to-noise ratio (S/N) of
4.5~pixel$^{-1}$ for objects with a $g$-band magnitude of 20.2. The
spectra are flux- and wavelength-calibrated, with 4096 pixels from
3800 to 9200~\AA\ at a resolution $R \equiv \lambda/\Delta \lambda
\approx 1800$. Redshifts are determined with pipeline code ({\tt
spectro1d}) by either cross-correlating the spectra with templates or
measuring emission lines; for galaxies, the typical redshift
accuracy is about $30$ \kms. {\tt spectro1d} also automatically classifies
the spectra into quasars, galaxies, stars, or unknown.

We correct the SDSS spectra for Galactic extinction using the
extinction map of Schlegel et~al. (1998) and the reddening curve of
Fitzpatrick (1999), and initially transform them into the rest frame
using the redshifts provided by the SDSS pipeline.

\subsection{Continuum and Emission-line Fitting}
%% a: zhou06; b: agn-dominated, Dong 08; c:lineprofile fitting, Dong 05
We summarize the procedures to model the continua and
emission-line profiles, as well as several updates to our previous
treatments. The fits are implemented with Interactive Data
Language (IDL) and based on the MPFIT package (Markwardt 2009),
which performs \chisq-minimization using the Levenberg--Marquardt
technique.

%%% #1: for agn--host composite spectra
Taken within a 3\arcsec-diameter aperture, many SDSS spectra have
a significant contribution from host galaxy starlight. Careful removal
of starlight, especially stellar absorption features, is essential
for the detection and accurate measurement of possible AGN emission lines
(see Ho et~al. 1997a and Ho 2004 for detailed discussions). For this
purpose, a set of algorithms has been developed; we refer readers
to Zhou et~al. (2006) for details, and only stress here two key
issues with updates.
\begin{itemize}
    \item
Host galaxy starlight is modeled with six synthesized galaxy
spectral templates, which were built from the library of simple
stellar populations of Bruzual \& Charlot (2003) using the Ensemble
Learning Independent Component Analysis algorithm (Lu et~al. 2006).
The templates are broadened by convolution with a Gaussian to match
the stellar velocity dispersion of the host galaxy. Particularly, to
account for possible errors in the redshifts provided by the SDSS
pipeline, we loop possible redshifts near the SDSS value by shifting
the starlight model in adaptive steps; the fit with minimum reduced
$\chi^2$ is adopted as the final result. This is important to the
measurement of host galaxy velocity dispersion and weak emission
lines.
    \item
The optical \feii\ emission is modeled with two separate sets of
templates in \emph{analytical} form\footnote{The IDL implementation of the two
template functions is available at
http://staff.ustc.edu.cn/\~{ }xbdong/Data\_Release/FeII/Template/ \,.},
one for the broad-line system and the other for the narrow-line
system. These two sets of templates are constructed from
measurements of I\,Zw\,1 by V\'eron-Cetty et~al. (2004), as listed
in their Tables~A1 and A2; see Dong et~al. (2008) for details, and
Dong et~al. (2011) for tests of the \feii\ modeling. The analytical
form enables us to fit \feii\ multiplets of any width. This is
especially useful in the case of low-mass AGNs, which often have
\feii\ lines significantly narrower than those of I\,Zw\,1, from
which almost all existing empirical \feii\ templates have been built
(cf. Greene \& Ho 2007b).
\end{itemize}

During the fitting two kinds of spectral regions had been masked
out: bad pixels as flagged by the SDSS pipeline and wavelength
ranges that are seriously affected by prominent emission lines.

%% #2: pure-AGN spectra
In some AGN-dominated spectra where the starlight contribution is
insignificant, the broad lines are so strong and broad that most of the
continuum and \feii\ regions would be masked
out as emission-line regions using the above continuum-fitting
method (see Zhou et~al. 2006). In practice such AGN-dominated sources
have undetectable \caii\,K (3934~\AA), \caii\,H + H$\epsilon$ (3970~\AA), and
H$\delta$ (4102~\AA) absorption features (see the Appendix of Dong et~al. 2011).
For these cases we fit simultaneously the nuclear continuum, the
\feii\ multiplets, and other emission lines (see Dong et~al. 2008 for
details). We recalculate the reduced $\chi^2$ of the fits around
the \ha\ and \hb\ regions. Spectra with the reduced $\chi^2 > 1.1$
in either region are subjected for further refined fitting of the
line profiles.

%% #3: emission-line profiles
The emission-line profiles, particularly the \ha\ + \nii\ complex
(or \ha\ + \nii\ + \sii\ if \ha\ is very broad), are fitted using
the code described in detail in Dong et~al. (2005). We fit the
emission lines using various schemes, and the one with the minimum
reduced $\chi^2$ is adopted as the final result. Basically, each
emission line (narrow or broad) is fitted incrementally with as many
Gaussians as statistically justified. The statistical criteria we adopt for
convergence (i.e., tolerance threshold) is that either the reduced $\chi^2
\leq 1.1$ \emph{or} the fit cannot be improved significantly by
adding in one more Gaussian with a chance
probability less than 0.05 according to $F$-test.%
\footnote{\label{fn-ftest} %
We have found through experimentation that these criteria based on
the $\chi^2$-test and $F$-test work well, although theoretically these
goodness-of-fit tests holds only for linear models (cf. Lupton 1993;
see also Hao et~al. 2005).} %
We show some examples of the fits in Figure~\ref{fig-demo4}.

\subsection{Error Assessment for Emission-line Parameters}
The parameters of the emission lines, both narrow and broad, are
measured from their best-fitting multi-Gaussian models. For broad
lines, the measurement uncertainties of the line parameters arise from
statistical noise, continuum subtraction (i.e., starlight,
AGN (pseudo-)continuum, or both), and subtraction of nearby
narrow lines. Formally, the total error is the quadrature sum of the
three independent terms,
\begin{equation}\label{formal_err_eq}
\sigma^2_{\rm total} = \sigma^2_{\rm n} + \sigma^2_{\rm cont\_sub} +
\sigma^2_{\rm NL\_sub}~.
\end{equation}
The measurement uncertainty due to statistical noise,
$\sigma_{\rm n}$, is given by the fitting code
MPFIT.%
\footnote{~The random error of the flux of emission lines modeled
with two or more Gaussians is given by the code through
well-constructed model parameterization; e.g., for a 2-Gaussian
model, $f(\lambda)$, we can parameterize it as follows:
\begin{eqnarray}
f(\lambda,~ [\lambda_0^1,~w^1,~ f_{\rm total},~%
r_{\scriptscriptstyle \lambda},~r_w,~ r_{\scriptscriptstyle f}] ) ~=~%
G(\lambda,~[\lambda_0^1,~w^1,~ (1-r_{\scriptscriptstyle f}) %
f_{\rm total}] ) ~+~ G(\lambda,~%
[r_{\scriptscriptstyle \lambda} \lambda_0^1,~ r_w w^1,~%
r_{\scriptscriptstyle f} f_{\rm total}] ) ~,
\end{eqnarray}
where $G$ denotes Gaussian function and $f_{\rm total}$ is the line
flux. MPFIT gives the error for $f_{\rm total}$ directly.} %
In practice, the error term caused by possible mismatch of the
continuum models is hard to estimate for every spectrum.
For each of our low-mass AGN spectra, we visually inspect the
continuum subtraction to guarantee that the subtraction uncertainty
is at least much below the 1 $\sigma$ spectral flux density error;
in particular, we carefully check the higher-order Balmer absorption
lines in the cases where there is considerable contribution from
intermediate-aged stellar populations. On the other hand, in
low-mass AGN spectra the broad Balmer lines are relatively narrow,
and, just like the narrow lines, their measurement is little
affected by the placement of the large-scale continuum (see Section
2.5 of Dong et~al. 2008). This is particularly true for broad \ha\
emission, because for intermediate-aged stellar populations the \ha\
absorption feature is generally much weaker than \hb\ and other
higher-order lines (e.g., up to H$\epsilon$). Hence, we believe that
the $\sigma^2_{\rm cont\_sub}$ term has been well minimized and is
negligible as far as the broad \ha\ lines are concerned
in this study (cf. Footnote~\ref{fn-bhaflux-limit} below). %

As for the error term caused by the narrow line subtraction, its
relative significance depends on the width of the broad \ha\ line.
In the case of very broad \ha, so broad that in the line profile
inflections can be apparently seen in between the broad component
and surrounding narrow lines, this term should be small and
negligible. For a broad \ha\ line that is both narrow and weak,
this term may be significant, as the line deblending is highly
dependent on the narrow-line model. We estimate it as follows. As
described in Section~3.2 below, in the refined line-fitting stage we
have four sets of fitting results out of the four schemes that adopt
different narrow-line models. By using the $F$-test, we pick up the $n$
sets of fitting results that are worse than the best-fitting
set with a chance probability greater than 0.1, and then calculate
the error term for parameter $p$ as
\begin{equation}\label{eq_nl_sub_err}
\sigma_{\rm NL\_sub} = \sqrt{\frac{ \sum_{i=1}^{n} (p_i - p_{\rm
best})^2}{n}}~.
\end{equation}

Our analysis indicates that the mean statistical relative error for the broad
\ha\ fluxes is 4\%, and that the mean total relative error
is 20\%. For the FWHM of broad \ha, the mean statistical relative
error is 7\%, and the total relative error is 28\%. In what follows, when we
speak of the significance of the detection of a broad \ha\ line
(say, at the 3 $\sigma$ level), we mean its relative flux with
respect to the statistical error ($\sigma_{\rm n}$); when we refer to
the S/N of broad \ha, we mean the ratio of its flux to the total
error ($\sigma_{\rm total}$).

\section{Sample Construction}

We start with the 451,000 spectra classified by the SDSS spectroscopic pipeline
as ``galaxy'' or ``QSO'' at $z < 0.35$ in the SDSS
DR4 (444,465 ``galaxies'' and 6,535 ``QSOs'', with duplicate
observations counted). They were arrayed on 1052 plates of 640
fibers each, with a total spectroscopic sky coverage of 4783~deg$^2$
(Adelman-McCarthy et~al. 2006). We set the redshift limit so that
the \ha\ line lies within the SDSS spectral coverage. To all the
spectra we performed the continuum and emission-line fittings with
the methods described in Section 2. Based on the fitting results, we
first compile a parent sample of 8862 broad-line (type~1)
AGNs according to our carefully defined criteria for detecting broad \ha\
emission (Section 3.1), by using an
automated selection procedure (Section 3.2). We then
estimate the BH mass from the luminosity and width of the broad
\ha\ line using the formalism developed by Greene \& Ho (2007b).
Applying a high mass cut-off at $\mbh = 2\times 10^6$ \msun, we end up with
a low-mass BH sample consisting of 309 sources.

In statistical studies of this nature, it is important to consider the impact
of selection effects.  Optimally, sample selection should be
objective and clearly defined with quantitative criteria that are
sufficiently reliable to select {\it bona fide} sources and,
at the same time, sufficiently efficient to make the sample as complete as
possible. Moreover, the criteria should be designed in such a way
that it is easy to implement---usually through Monte Carlo
simulations---corrections for selection effects (see, e.g., Hao
et al. 2005; Greene \& Ho 2007a; Lu et al. 2010). In addition to having
well-defined and robust criteria, in practice an efficient, automated
selection procedure is also required to handle large-volume data sets such as
that from SDSS.  We have made a concerted effort to
design an efficient and effective selection procedure for these purposes.

\subsection{Criteria for Detecting Broad \ha\ Emission}
The broad \ha\ line is generally the strongest broad line in the
optical spectra of AGNs. When it is weak, as expected in low-mass
AGNs with low accretion rates, defining the criteria for broad \ha\
detection is by no means trivial (see, e.g., Ho et~al. 1997b; Greene
\& Ho 2004; Dong et~al. 2005; Hao et~al. 2005; Zhou et~al. 2006;
Greene \& Ho 2007a, 2007b). For the purpose of comparison, we
summarize some of the criteria adopted previously in the literature.
In the early efforts to mine broad-line AGNs from the SDSS Early Data
Release, Dong et~al. (2005) used a simple criterion for the detection
of broad \ha, by requiring that its S/N be $\geq 5$. Later, Zhou et~al. (2006)
used a stricter criterion, requiring the S/N of broad \ha\ to be $\geq 10$ to
select narrow-line Seyfert 1s.
%GH04, GH07a,b
Greene \& Ho (2007a) took a two-step procedure to select broad-line
AGNs (see also Greene \& Ho [2004] for a qualitative description of an
earlier version of their criteria). In the first step, they employed an
initial selection algorithm to select candidates with excess rms
deviation in the \ha\ region. In the second step, they performed
detailed line-profile fitting of the \ha~+~\nii\ region, and set the
broad \ha\ criteria as,\\
\indent (1) $\chi_{\rm N}^2 /\chi_{\rm B}^2 ~>~ 1.2 $, \\
\indent (2) Flux(total \ha) /\,rms $~>~ 200$ \AA, and \\
\indent (3) EW(total \ha) $~>~ 15$ \AA,\\
where $\chi_{\rm B}^2$ and $\chi_{\rm N}^2$ represent the \chisq\ of the fits
with and without accounting for broad \ha, respectively, and rms
represents the rms deviations in the $6400-6700$~\AA\ region of the
continuum-subtracted spectra. Criterion 1 requires that
adding a broad \ha\ component to the fit results in a 20\% decrease in
\chisq; this empirical rule is inspired by $F$-test statistics used to
evaluate the significance of line fits (Hao
et~al. 2005).  %
Criteria 2 and 3 are aggressive cuts (their ``detection threshold'')
to guarantee the reliability of BH mass estimates based on broad \ha\
line width and luminosity. As the detection threshold is quite
strict, the low-mass ($\mbh < 2 \times 10^6$~\msun) AGN sample tends to have
rather high Eddington ratios ($\lratio > 0.1$
mostly; see Figure~1 of Greene \& Ho 2007b).  In an effort to find more
low-$\lratio$ objects, Greene \& Ho (2007b) manually selected additional
sources and tagged them as candidate low-mass systems (their $c$ subsample).

In this work, we select broad-line AGNs based directly on the results of the
emission-line fits.  Our broad \ha\ criteria are the following: \\
\indent (1) $P_{F-{\rm test}} < 0.05$, \\
\indent (2) Flux(\bha) $> 10^{-16}$ \flux, \\
\indent (3) S/N(\bha) $\geq 5$, and \\
\indent (4) $h_{\rm B} \geq 2$\,rms, \\
where $P_{F-{\rm test}}$ is the chance probability given by the $F$-test
that adding a broad \ha\ in the model is significant; %
S/N(\bha) $=$ Flux(\bha)$/\sigma_{\rm total}$, as stated in Section
2.3; $h_{\rm B}$ is the height of the best-fit broad \ha\ line; %
and rms is the rms deviation of the continuum-subtracted spectra in the
emission-line--free region near \ha. %
Criterion 1 gives the statistical significance of detecting
broad \ha\ given by $F$-test (see Footnote~\ref{fn-ftest}). The
lower limit of the broad \ha\ flux in Criterion 2 is set to
eliminate possible spurious detections caused by systematic errors,
such as those caused by inappropriate continuum subtraction (cf.
Footnote~\ref{fn-bhaflux-limit}). Criterion 3 is to ensure the
reliability of broad \ha\ in terms of S/N. Criterion 4
minimizes spurious detections mimicked by narrow-line wings or any
large-scale fluctuation in the continuum; it is set by trial-and-error
based on the confirmation of the AGN nature in some of the objects
by the presence of broad \hb\,%
\footnote{An example of this kind is SDSS J120216.04$+$060937.2,
which has $h_{\rm B}$ of broad \ha\ close to 2~rms but yet the broad
components of both \ha\ and \hb\ have S/N $>5$. We note that, for a
few low-mass AGNs having extremely strong narrow lines but weak
broad lines, broad \hb\ can be measured more reliably than broad
\ha. See also the case of POX~52 in Barth et~al. (2004).} or by our
long-slit observations using the MMT 6.5-m telescope with a narrow
slit width under good seeing conditions. Our final results are not
very sensitive to the exact value chosen for the $h_{\rm B}$
threshold; changing the value by $\pm$10\% has little impact on
final results.  The S/N threshold in Criterion 3 generally selects
objects with $h_{\rm B}$ much larger than 2~rms.

\subsection{The Automated Selection Procedure}
%%WM: \com{there are some places repeating what are to be said or already
%%said, which I removed. also i made the context more concise but the
%%the key ingredients kept...} The automated selection procedure is
%% composed of two stages.
%an initial line-fitting stage and a refined one.
The selection procedure is composed of two stages. In the initial
fitting stage, each spectrum is fitted quickly, with the aim of
reducing the number of broad \ha\ candidates passed to the second
stage. Next, the \ha~$+$~\nii\ complex (or
\ha~$+$~\nii~$+$~\sii\ if \ha\ is very broad) is fitted carefully to
get reliable parameters for broad \ha.  We then select {\it bona fide} broad
\ha\ emitters based on the criteria outlined above.
We describe these two stages in detail; a flow chart is given
in Figure~\ref{fig-flowchart}.

%% step 1 of the initial fitting
The initial stage has three steps. The first is to
build a narrow-line model from the \sii\,$\lambda\lambda6716,6731$
doublet (or the core component of the
\oiii\,$\lambda\lambda4959,5007$ doublet if \sii\ is weak).
%The \sii\ lines are assumed to have the same redshift and profile.
The two \sii\ lines are assumed to have the same
profile and are fixed in separation by their laboratory wavelengths.
Each line is fitted with one Gaussian, and, if
either is detected at S/N $\geq 10$, more Gaussians are added.
 The final fit is achieved
following the tolerance threshold as stated in Section 2.2.
The \oiii\ doublet is fitted in a similar way.%
\footnote{To account for the possible effect of the broad \ha\ and
\hb\ lines (mimicking a local ``pseudo-continuum'') on the fitting
of \sii\ and \oiii, respectively, a local 1st-order polynomial is
added to the model at this stage.} %
The flux ratio of the doublet $\lambda$5007/$\lambda$4959 is fixed
to the theoretical value of 2.98. The \oiii\ profile usually shows a
broad wing blueward of the line core (e.g., Greene \& Ho 2005a;
Zhang et al. 2011); thus, it is fitted with
multiple Gaussians, one for the line core, and one or more for the wings.
In this way we build a model for the narrow-line profile, using the
best-fit model for \sii\ if we can, or the core of \oiii\ if the \sii\
lines are too weak ($<5$ $\sigma$ significance).

%% step 2 of the initial stage
Step 2 of the first stage is to fit the \ha~+~\nii\ region. The
narrow \ha\ component and the \nii\,$\lambda\lambda6548,6583$
doublet are modeled with the narrow-line model obtained above,
assuming that these three lines have the same redshift and profile.
The flux ratio of the \nii\ doublet $\lambda$6583/$\lambda$6548 is
fixed to the theoretical value of 2.96. An additional Gaussian to
account for possible broad \ha\ is added if the \chisq\ decreases
significantly with a $F$-test probability $<$0.05. At this point the
centroid of the broad component is fixed to that of narrow \ha,
while the width and flux are left as free parameters.

In Step 3, a reduced number of objects with candidate broad \ha\ are
selected, which will be passed to the next stage. They are selected
from those with a possible ``broad'' \ha\ component determined from Step 2,
which have %
(1) FWHM of the broad component greater than that of any
narrow lines (particularly \oiii\,$\lambda5007$); and
(2) flux 3 times greater than the flux
error given by MPFIT (namely $\sigma_{\rm n}$), and meanwhile
greater than $10^{-16}$ \flux. These criteria are physically
meaningful and practical (see Section~3.1); in particular, we
believe from our experiments that the sensitivity of detecting a
broad \ha\ line in the general SDSS database is greater than %%no deeper than
$\sim 10^{-16}$ \flux.\,%
\footnote{\label{fn-bhaflux-limit}%
We consider ``broad'' \ha\ components detected at a flux level
$<10^{-16}$ \flux\ in the initial fitting stage to be mostly spurious,
as at such a level the line detection is highly susceptible to
(even slight) uncertainty in the continuum subtraction. For
instance, a slight offset in the continuum level of $5 \times
10^{-18}$ erg~s$^{-1}$~cm$^{-2}$~\AA$^{-1}$, comparable to or
smaller than the 1 $\sigma$ measurement error around \ha\ in
SDSS spectra, would result in a spurious broad \ha\ detection with
such a flux. In fact, our final broad-line sample has a minimum
broad \ha\ flux of $(5.2 \pm 0.5) \times 10^{-16}$ \flux. On the
other hand, this also means that an uncertainty at the 1 $\sigma$
level in the continuum subtraction would introduce an error of at most 20\%
to the broad \ha\ flux, even when the line is extremely weak.} %

The initial stage of line fitting produced $\sim$49,600 spectra (11\%) that
appear to have a ``broad'' \ha\ component that is statistically significant
according to the $F$-test (Step 2), of which about 23,700 are regarded
as physically reasonable candidates (after Step 3). These candidates
are passed to the second stage for refined fitting of their
emission-line profiles.

%%%%% the 2nd stage: refined fitting the em-line profiles
The final fits are performed using different schemes for modeling the
narrow \ha\ line. Since narrow \ha\ (and \hb) line can arise from emitting
regions with a larger range of density and ionization than forbidden
lines such as \nii\ and \sii, it may have a different profile than
\sii\ (see Section 2 of Ho et~al. 1997b for details; also Zhang
et~al. 2008). Considering this fact, we employ four different models
for treating narrow \ha: %
(1) a single-Gaussian model built from narrow \hb, if broad \hb\ is
not detected in the first stage \emph{and} narrow \hb\ can be fitted
well with one Gaussian; %
(2) a model built from the best-fit \sii\ with one or more Gaussians,
as described above; %
(3) a single-Gaussian model from the best-fit core component of \oiii; %
(4) a multiple-Gaussian model from the best-fit global profile of \oiii.
In each scheme, the \ha~$+$~\nii\ region is fitted in a similar way
as in Step 2 of the first stage, except that the possible broad \ha\
is now modeled with as many Gaussians as statistically justified (see
Section 2.2), and the centroid of broad \ha\ is no longer fixed.
The fitting scheme that gives the smallest reduced $\chi^2$ is
adopted as the final result. The {\em bona fide} broad-line AGNs are
then selected based on the broad \ha\ criteria described in
Section~3.1.

Our analysis yielded a total of 8862 sources (with duplicates removed) with
secure detections of broad \ha.  These form our parent sample of broad-line
AGNs (namely Seyfert 1s and quasars).

\subsection{Low-mass AGN Sample}
It has become possible since the last decade to estimate BH masses
in type 1 AGNs using single-epoch spectra, thanks to the significant
advances in the reverberation mapping experiments of nearby Seyfert
galaxies and quasars (e.g., Kaspi et al. 2000; Peterson et al.
2004). In this paper, we adopt the \mbh\ formalism presented by
Greene \& Ho (2007b), which makes use of the luminosity and FWHM of
the broad \ha\ line (see their equation A1). This formalism is based
on the radius-luminosity relation reported by Bentz et~al. (2006) and assumes
a spherical broad-line region (BLR) with a virial coefficient of $f=0.75$. For
ease of comparison, we simply
adopt the same upper limit on BH mass, $2 \times 10^6 \msun$, which
Greene \& Ho (2007b) used to define low-mass BHs.  This threshold was motivated
by the smallest nuclear BH that has been reliably measured dynamically, $\mbh
= 2.5 \times 10^6 \msun$ in M32 (Verolme et~al. 2002).
This cut produces a final sample of 309 AGNs with low-mass BHs.

Note that Wang et~al. (2009) re-calibrated the BH mass formalisms
based on single-epoch broad \hb\ and \mgii\,$\lambda 2800$ lines,
stressing the nonlinear relation between the virial velocity of the
BLR clouds and the FWHMs of the single-epoch broad lines. This
nonlinearity probably arises from the mixture of non-virial
components in the total profile of broad emission lines
in single-epoch spectra (see Section~4.2 of Wang et~al. 2009 for
detailed discussion, and also Collin et~al. 2006 and Sulentic et~al.
2006). The mass estimators of Wang et~al. (2009), however, were
calibrated using reverberation mapping data for AGNs with BHs in the mass
range $10^7$ to $10^9$~\msun, which does not cover the low-mass
regime studied in this paper. Considering the uncertainty in
extrapolating the empirical relations to the lower \mbh\ end by more than an
order of magnitude, as well as the ease of comparison with previous
samples, we adopt the same mass formalism as used by Greene \& Ho (2007b).
Moreover, we point out that the non-virial components, if any,
may be less significant in our sample of low-mass objects, whose broad
\ha\ lines tend to be roughly symmetrical and generally close to a
Gaussian. The statistical
uncertainty of the BH masses is 0.3~dex (1 $\sigma$;
cf. Wang et~al. 2009), which is dominated by the systematics of the
virial method rather than statistical measurement errors (see Xiao
et~al. 2011 for a detailed analysis).

We estimate their bolometric luminosities and Eddington ratios in
the same way as Greene \& Ho (2007b). That is, we adopt $L_{\rm bol}
= 9.8~\lambda L_{\lambda} (5100$~\AA) (McLure \& Dunlop 2004), which
is consistent with the broad-band SED of NGC\,4395 (Moran et~al.
1999), while $\lambda L_{\lambda} (5100$~\AA) is calculated from the
\ha\ luminosity (Greene \& Ho 2005b).
The Eddington ratio (\lratio) is the ratios between the bolometric and
Eddington luminosities.
The Eddington luminosity (\ledd) is the maximum luminosity
of the central BH powered by spherical accretion,
at which the gravity acting on an electron--proton pair
is balanced by the radiation pressure due to electron Thomson scattering;
$\ledd  = 1.26 \times 10^{38}$ (\mbh/\msun) erg~s$^{-1}$.
Table~1 shows the basic
properties of the sample; Table~2 lists measurements of the
emission-line parameters; and Table~3 gives the luminosities and BH
masses. The emission-line parameters are calculated from the
best-fit models of the line profiles. The flux of the
\feii\,$\lambda4570$ emission blend is integrated in the rest-frame
wavelength range 4434--4684~\AA. For all measured emission-line
fluxes, we regard the values as reliable detections if they have
greater than 3 $\sigma$ significance; otherwise, we adopt the
3 $\sigma$ error as an upper limit.
The data and fitting parameters for the low-mass BH sample
are available online for the
decomposed spectral components (continuum, \feii, and other emission lines).
\footnote{Available at \\
http://staff.ustc.edu.cn/\~{}xbdong/Data\_Release/IMBH\_DR4/\,.
%, together with auxiliary code to explain the parameters and to
%demonstrate the fitting.
At present the basic parameters for the
parent AGN sample are available only on request.}

Figure~\ref{fig-our4dist}
shows the distributions of the luminosity and FWHM of broad \ha,
\mbh, and \lratio\ for the low-mass BH sample, overplotted on contours of
the parent sample of 8862 broad-line AGNs.
We note that, in the \mbh--\lratio\ plane (left panel),
the lack of AGNs in the large-\mbh,
large-\lratio\ region (upper right corner) reflects cosmic
downsizing of supermassive BH activity (see, e.g., Heckman et~al.
2004), in the sense that toward lower redshifts AGN activity gets shifted from
massive BHs to their lower-mass counterparts.
The dearth of objects in the small-\mbh, small-\lratio\
region (lower left corner) reflects the selection effect due to
host galaxy starlight contamination (i.e., the difficulty in
detecting broad \ha; see Section~3). The lone outlier is SDSS
J103234.85$+$650227.9 (NGC~3259), which has a very low redshift
($z=0.0057$). Its broad \ha\ line was first revealed from a high-resolution,
high-S/N spectrum taken with the 10-m Keck telescope; the
central stellar velocity dispersion of the host galaxy is $43 \pm 4$ \kms\
(Barth et~al. 2008).

\subsection{Comparison with the Greene \& Ho (2007a, 2007b) Samples}

Since our study represents an independent effort to select low-mass AGNs
from what is otherwise identical SDSS data, it is worthwhile to compare our
sample with that of Greene \& Ho (2007b). Of the 309 objects
in our sample, 160 are not included in Greene \& Ho
(2007b), 85 are not in their parent broad-line AGN sample
(Greene \& Ho 2007a), and 61 are not in either Greene \& Ho
(2007a) or Greene \& Ho (2007b). On the other hand, of the
229 objects in Greene \& Ho (2007b), 11 are not in the
official data set of SDSS DR4 (Fermilab version) but rather were taken
from the reductions of D.~Schlegel at Princeton (J.~Greene 2010,
private communications), 37 are not in our sample
due to our broad-line selection criteria, and an additional 32 are not in our
low-mass sample because the estimated \mbh\ exceeds the mass
cut. In total, there are 149 objects in common between the two studies
(65\% of the entire sample of Greene \& Ho [2007b] and 48\%
of this present sample).

In light of the considerable discrepancy in the number of objects
between the two samples, we compare their distributions of redshift,
broad \ha\ luminosity and FWHM, \mbh, and \lratio\
(Table~\ref{tab-compgh07}).  The \lratio\ distributions are shown in
Figure~\ref{fig-compgh07ell}. The distributions of redshift, broad
\ha\ luminosity and FWHM, and \mbh\ are similar (within 0.1~dex on
average) between our sample and the entire low-mass BH sample of
Greene \& Ho (2007b; including the 55 candidates below their
detection threshold that dominate the $\lratio <0.1$ population).
The median \lratio\ of the present sample is 0.2~dex lower than that
of their entire sample.  However, we notice that the \lratio\
distribution of their entire sample is not smooth.  The \lratio\
distribution of their uniformly selected sample of 174 objects,
however, is smoother; this suggests that manual selection may have
introduced inhomogeneities into their sample (cf. Xiao et~al. 2011).
When only the two uniformly selected samples are considered, there
is a more significant difference in the \lratio\ distribution: our
sample has a smaller median value ($-0.7$~dex vs. $-0.4$~dex,
computed in the logarithm) and a larger standard deviation
($0.5$~dex vs. $0.3$~dex) compared to Greene \& Ho (2007b).  The
\mbh\ distributions of the two uniformly selected samples are
similar. Hence the difference in \lratio\ is due to fainter broad
\ha\ emitters detected in our sample; it has more objects in the
low-luminosity end, with a median luminosity of broad \ha\ 0.3~dex
lower than their uniformly selected sample (see
Table~\ref{tab-compgh07} and Figure~\ref{fig-mags}). %
While there are 17 (10\%) objects with $\lratio \leqslant 0.1$ in the
uniformly selected sample of Greene \& Ho (2007b) (and 59 in total
in their entire sample), 89 (30\%) objects of our sample have
$\lratio \leqslant 0.1$. Thus, our approach has the advantage of
systematically selecting more low-mass AGNs at low accretion rates.
% the same bolometric corrector: L_5100 *9.8, June 2011, xbdong

Since the spectral fitting methods adopted in this work are, in detail,
different from those of Greene \& Ho (2007b), we also compare the
emission-line parameters derived from the two samples,
using the 149 objects in common (Figure~\ref{fig-common149}). The mean
and standard deviation of the difference in the luminosity of broad
\ha\ (computed in logarithm) are 0.15~dex and 0.14~dex,
respectively; for the FWHM of broad \ha, they are
0.02~dex and 0.09~dex, respectively.
On average, the emission-line quantities of the two
samples are consistent within 1 $\sigma$ dispersion (about
0.1~dex).  The only noticeable exception is the broad \ha\ luminosity, for
which the values of Greene \& Ho (2007b) are systematically higher than ours
by 0.15~dex.  After consulting with D.~Schlegel (2011, private communications),
we suspect that this systematic offset can be traced to differences in the
calibration methods between the officially released SDSS data set we used and
the Princeton version used by Greene \& Ho (2007b).  In any case, this
calibration discrepancy is smaller than the difference between the two samples.

%XX New subsection on MMT observations

\subsection{Supporting New Observations}

%%% 52652-0945-172, J1005+5432 MMT spec, in the brand-new subset
%
We have an ongoing program to observe a subsample of the nearby ($z<0.1$)
low-mass BHs with the 6.5-m MMT
telescope, using a 1\arcsec-wide slit under good seeing
conditions. Compared to the SDSS spectra taken through 3\arcsec-diameter
fibers, the contamination from the host galaxy starlight is
significantly reduced in the MMT spectra. One of our goals is
to check independently the reliability of the broad lines,
particularly for the new objects that were not included in Greene \&
Ho (2007a, 2007b). The SDSS spectra of four such objects are
shown in Figure~\ref{fig-demo4}; see also N.~Jiang et~al. (in
preparation; their Figure~1) for SDSS\,J140040.56$-$015518.2
(UM~625; $\lratio=0.04$), which has extremely strong narrow emission
lines, similar to NGC~4395 and POX~52. Here we show in
Figure~\ref{fig-mmt} the MMT spectrum of yet another object,
SDSS\,J100510.51$+$543255.5. The broad H$\alpha$ component matches
very well the one in the SDSS spectrum (see the right inset).
Moreover, in the original MMT spectrum the \hb\ line already shows a
profile significantly broader than the \oiii\,$\lambda4959,5007$
doublet (see the left inset), even without subtracting the
starlight. Hence the broad \hb\ component is real, definitely not an
artifact caused by overestimating the \hb\ absorption in the
starlight spectrum.
%% MMT spectra end; below Chandra observations.
As another independent check, we also obtained {\it Chandra}\ X-ray
observations for four nearby objects with $\lratio < 0.1$. We find
that their X-ray emission is indeed weak up to 10\,keV, consistent
with the low \lratio\ nature inferred from their optical spectra
(W.~Yuan et~al., in preparation).

%%%%%%%%% Section 4 %%%
\section{Sample Properties}

As described in last section, the present sample and the previous one by
Greene \& Ho (2007b) share only about half of the objects in
common. Moreover, the new sample spans a wider regime in the
parameter space by including many more low-accretion systems. It is
therefore of particular interest to investigate the ensemble
properties of this new sample. In this initial paper we present some
general results pertaining to the properties of the AGNs
and the host galaxies based on the SDSS data. We defer studies of
other properties, such as the multi-wavelength properties, to future
work.

\subsection{Emission Lines}

We begin by exploring the distribution of the low-mass BH sample
in the diagnostic diagrams of narrow-line ratios
(Figure~\ref{fig-bpt}), which are a powerful tool to separate
Seyfert galaxies, low-ionization nuclear emission-line region
sources (LINERs; Heckman 1980), and \hii\ galaxies (Baldwin et~al.
1981; Veilleux \& Osterbrock 1987; Ho et~al. 1997a; Kewley et~al.
2001; Kauffmann et~al. 2003b; Kewley et~al. 2006). About two-thirds
of the objects are located in the conventional region of Seyfert
galaxies, in terms of either the empirical demarcation line of
Kauffmann et~al. (2003b; the dashed line in panel \textit{a} of
Figure~\ref{fig-bpt}) in the \oiii~$\lambda5007$/\hb\ versus
\nii~$\lambda6583$/\ha\ diagram, or the empirical lines of Kewley
et~al. (2006; the dotted lines in panels \textit{b} and \textit{c}) in
the \oiii~$\lambda5007$/\hb\ versus \sii~$\lambda\lambda6716,6731$/\ha\ and
\oi~$\lambda6300$/\ha\ diagrams.
The remaining one-third of the objects are located in the region for
\hii\ galaxies in the latter two diagrams in terms of the empirical
demarcation lines of Kewley et~al. (2006), or located mostly in the
region of the so-called transition objects between \hii\ galaxies
and Seyfert galaxies in the \oiii~$\lambda5007$/\hb\ versus
\nii~$\lambda6583$/\ha\ diagram in terms of the theoretical
maximum starburst line of Kewley et~al. (2001; the dotted line in
panel \textit{a}). From visual inspection of their spectra and
images, we suspect that the narrow-line \hii\ characteristic of these
objects is mainly caused by the inclusion in the SDSS fiber aperture
of emission from star formation regions in the host galaxies. %
Only a few ($\sim10$) of the 309 objects show a LINER characteristic
in the three diagrams (lower right region of each panel), according
to the Seyfert--LINER demarcation lines of either
\oiii~$\lambda5007$/\hb\ $= 3$ (see panel \textit{a}) or
Kewley et~al. (2006; see panels \textit{b} and \textit{c}). %
%%\xbcomm{To comment on the few LINERs. --7/17/2011.}
The Eddington ratios of the type 1 LINERs range from 0.01 to 1.2,
with a median of 0.07 and a standard deviation of 0.43. Given the
large measurement uncertainty of the Eddington ratios, this is
broadly consistent with the notion that LINERs have low accretion
rates (e.g., Ho 2004; Kewley et~al. 2006; Ho 2009).

%% broad-line begins, --xbdong, June 4, 2011
With accurate flux measurements of the broad \ha,
\feii\,$\lambda4570$, and \oiii\,$\lambda5007$ emission lines, it
would be interesting to explore for the first time the emission-line
properties of low-mass AGNs in the framework of the
eigenvector 1 AGN parameter space of Boroson \& Green (1992; EV1 or PC1).
EV1 is dominated by the anticorrelation between the
strengths of \feii\,$\lambda4570$ and \oiii\,$\lambda5007$, and is
suggested to be physically driven by the relative accretion rate or
the Eddington ratio (Boroson \& Green 1992; Marziani et~al. 2001;
Boroson 2002; cf. Dong et~al. 2011; Zhang et~al. 2011). In a
detailed re-analysis of this issue using a large, homogenous AGN
sample, Dong et~al. (2011) found that the intensity ratio of
\feii\,$\lambda4570$ to \oiii\,$\lambda5007$ indeed correlates most
strongly with \lratio, whereas the apparent correlations with broad-line width,
luminosity, and \mbh\ are only a secondary effect. Surprisingly,
however, Dong et~al. (2011) also found that the correlation of
\feii/\oiii\ with \lratio\ is not as strong as that of the EW of
\feii\,$\lambda4570$ itself, and actually much weaker than that of
the intensity ratio of \feii\,$\lambda4570$ to other broad lines
(such as broad \hb\ and \mgii\,$\lambda2800$). %
We carry out similar analysis for our low-mass AGN sample. Here we
use broad \ha\ instead of broad \hb, and adopt the generalized
Spearman rank correlation test to account for censored data.
It turns out that the correlations of both \feii/\oiii\ and
\feii/\bha\ with \lratio\ are much stronger than those with
FWHM(\ha$^{\rm B}$), \ha$^{\rm B}$ luminosity, or \mbh; this is just
as expected, given the restricted ranges of the latter three
quantities for the low-mass AGN sample. Moreover, both intensity
ratios have equally strong correlations with \lratio, with Spearman
coefficients $\rs = 0.51$ (a chance probability $\pnull = 7 \times
10^{-27}$). The same analysis is also performed using the parent
sample, taking advantage of its large dynamic ranges in both
\lratio\ and \mbh\ (Figure~\ref{fig-our4dist}). Again, both
intensity ratios have the strongest correlations with \lratio, and
the one involving \feii/\bha\ has a slightly higher correlation
coefficient ($\rs = 0.42$ vs. 0.37). The correlations with  \lratio\
are illustrated in Figure~\ref{fig-pc1}.

%\xbcomm{Todo: to add a new paragraph to briefly discuss about NLS1s,
%in terms of \lratio\ or EV1. --7/17/2011.}
Note that, since $\mbh \propto {\rm FWHM}^{2}$, most of the low-mass
AGNs in the sample would be classified as narrow-line Seyfert 1s
(NLS1s) based on their modest values of \ha\ FWHM.  In fact, some of
the objects are included in the SDSS NLS1
sample of Zhou et~al. (2006). NLS1s are observationally defined as
Seyfert 1s having broad \hb\ FWHM $\lesssim 2000$~\kms, often
characterized by strong \feii/\hb\ and weak \oiii/\hb\ in their
optical spectra (see, e.g., Laor 2000 and Komossa 2008 for reviews).
Their narrower broad lines are generally believed to arise from their
smaller BH masses compared to normal Seyfert nuclei, and their
multi-wavelength properties are found to lie at the extreme end of
EV1 associated with high accretion rates %
%%% better to use accretion rates here instead of Eddington ratio, due
%%% to the saturation of \lratio\ in slim disks (Luis).
(Boroson \& Green 1992; cf. Desroches et al. 2009 for a recent
discussion). Given that in this sample we have made efforts to find low-mass
BHs at low accretion rates, most of the objects do not accrete
close to the Eddington limit; the median \lratio\ is only 0.2 (see
Figures \ref{fig-our4dist} and \ref{fig-compgh07ell} and
Table~\ref{tab-compgh07}). We refer readers to Greene \& Ho (2007b;
their Section~3.2) and Ai et al. (2011; their Section~6.3) for
more detailed discussions on low-mass AGNs and their relationship to NLS1s.

\subsection{Properties of the Host Galaxies}

Since the discovery of massive BHs at the centers of galaxies, there
have been several long-standing puzzles.  Is there a
lower limit to the mass of central BHs?  What are the smallest
galaxies harboring central BHs?  What is the connection between the BH
and the host galaxy in the low-mass regime?
Studies on the present sample may contribute
to the understanding of these questions. Below we present some
preliminary results on the host galaxy properties based on the SDSS
imaging data.

\subsubsection{Luminosities}
%% firstly, the same investigation as GH07b, June 12, xbdong
%
We calculate the host galaxy luminosities in the same way as Greene
\& Ho (2007b). The contribution of the AGN light to the total
Petrosian $g$-band magnitude is subtracted by using the \ha\
luminosity to predict the contribution of the AGN continuum to the broad-band
photometric magnitude.  This is accomplished using the
$L_{\rm \ha}$--$L_{5100}$ relation of Greene \& Ho
(2005b) and assuming that the AGN continuum follows the shape
$f_{\lambda} \propto \lambda^{-1.56}$
(Vanden~Berk et~al. 2001).  Because of the possibility of aperture losses,
we do not trust host galaxy estimates based on the starlight decomposition
of the spectra. $K$-corrections of the host galaxy magnitudes are
estimated using the routine of Blanton \& Roweis (2007). The
distributions of the AGN, host, and total luminosities are displayed
in Figure~\ref{fig-mags} (black); also plotted are the corresponding
distributions from  Greene \& Ho (2007b), both their uniformly selected
sample (red, solid) and their entire sample (red, dotted).  Compared to the
Greene \& Ho (2007b) sample, while the present sample has more
objects with low AGN luminosities, it contains, interestingly, more
objects with higher host galaxy luminosities.  This presumably reflects the
fact that we have put special effort to detect sources with weaker
emission lines, as a consequence of which we can tolerate objects with a
more significant host galaxy starlight contribution.
The present sample has a median AGN luminosity of $M_g =
-17.7$~mag and a median host galaxy luminosity of $M_g = -20.2$~mag,
0.7~mag fainter and 0.9~mag brighter, respectively, than the corresponding
medians of the uniformly selected sample of Greene \& Ho (2007b).
Almost all the host galaxies (304 of 309) are brighter
than their nuclei; their luminosity differences have a median of
2.5~mag and a standard deviation of 1.2~mag. The peak of the host
galaxy luminosity distribution is comparable to the characteristic
luminosity of $M_g^* = -20.1$~mag at $z=0.1$ (for our assumed
cosmology; Blanton et~al. 2003), and only 10 objects have host
galaxy luminosities $M_g > -18.0$~mag. Note that our method of
AGN--host separation tends to underestimate the AGN luminosity because of
fiber losses, and hence the host galaxy luminosity is generally
overestimated. According to the experiments of Greene \& Ho (2007b),
typically the systematic overestimation should be within 0.3~mag.
Another source of error comes from the conversion from \ha\
luminosity to $g$-band magnitude and the assumed continuum slope; it
is estimated to be $\sim 0.1$~mag (1 $\sigma$).

\subsubsection{Colors and Morphologies}
%%\xbcomm{To rewrite the following paragraph, and add the morphology
%%results. --7/17/2011.} %
%% 1: color method as GH07b %
Due to the short exposure time (54 seconds) and mediocre seeing
conditions ($\sim$1\farcs5) of the SDSS images, direct visual classification of
galaxy morphology is impossible for most of the objects in this sample, even
those at $z<0.05$. Although the AGN contribution is not large in many
cases, the central point source nonetheless significantly affects the
central structure of the host.  Disk features such
as spiral arms and rings are generally
smeared out, making it hard to distinguish
face-on disks from elliptical or spheroidal galaxies without images of
much higher resolution (Greene et al. 2008; Jiang et al. 2011). Following
Greene \& Ho (2007b), we try to ascertain some rough information from galaxy
colors, with the AGN contribution removed from the Petrosian
magnitudes. The $u-g$ colors have a mean value of 1.15 mag with a standard
deviation of 0.35 mag; the $g-r$ colors have a mean of 0.57 mag and a
standard deviation of 0.13 mag. According to Fukugita et~al. (1995), these
colors correspond to typical values of Sbc galaxies. %

\subsubsection{Stellar Populations}
%% D4000--Hd_A plane, July 24, 2011, xbdong
We briefly examine the stellar
populations of the host galaxies of our low-mass BHs using
two stellar indexes, the 4000~\AA\ break ($D_{4000}$) and the
rest-frame equivalent width of the H$\delta$ absorption line
(H$\delta_{\rm A}$). As discussed in Kauffmann et al. (2003a), $D_{4000}$ is
an excellent age indicator for young stellar populations ($D_{4000} <1.5$ for
ages $<1$~Gyr), while for
older populations it also depends strongly on metallicity.
H$\delta_{\rm A}$ also does not depend strongly on metallicity
except for old populations. Large H$\delta_{\rm A}$ values indicate
a burst of star formation that ended 0.1--1~Gyr ago. Altogether, the
locus of galaxies in the $D_{4000}$--H$\delta_{\rm A}$ plane
is a powerful
diagnostic of whether stars have been forming continuously
or in bursts over the past 1--2~Gyr. Galaxies with continuous star
formation histories occupy an intrinsically very narrow strip in the plane
(without taking into account measurement errors of the indices),
whereas recent starbursts have large H$\delta_{\rm
A}$ values, significantly displaced away from the locus of continuous
star formation histories (Kauffmann et~al. 2003a).

We take Kauffmann et al.'s (2003a) definition and calculation method of the
two indices. The calculation is based on the
spectrum of the decomposed starlight component, as described in
Section~2.2. For reliable measurement of the indices, we only use
the 262 objects that have AGN contribution less than 75\% at
4000~\AA\ in the SDSS spectra (cf. Zhou et al. 2006). For
comparison, we also calculate the two indexes for $\sim$318,500 inactive
galaxies in the SDSS DR4 that have S/N $> 5$~pixel$^{-1}$ around
4000~\AA.  For normal galaxies the typical 1 $\sigma$ errors on
$D_{4000}$ are 0.05, but for H$\delta_{\rm A}$ they are substantial, on the
order of 1.4~\AA\ (Kauffmann et al.~2003b).
For our low-mass AGN hosts, the measurement errors
should be even larger due to the additional uncertainty from the
AGN-host decomposition. We assign an uncertainty of 15\% to the
decomposition (Zhou et~al. 2006) and estimate total 1 $\sigma$
errors on $D_{4000}$ and H$\delta_{\rm A}$ to be 0.1 and 1.6~\AA,
respectively.

Figure~\ref{fig-d4k} shows the distribution of the 262 low-mass AGN
hosts in the $D_{4000}$--H$\delta_{\rm A}$ plane, superposed on contours
for the SDSS inactive galaxies. Note that the negative values of
H$\delta_{\rm A}$
are not caused by contamination from H$\delta$
emission, but due to the definition and corresponding calculation
method of the line index (see Kauffmann et~al. 2003a; Worthey \&
Ottaviani 1997). A few results can be inferred. First,
the majority of the low-mass AGN hosts (174 out of 262) have $D_{4000} <
1.5$---that is, they have mean stellar ages less than 1~Gyr. Second,
although the measurement errors are large, the distribution of the low-mass
BH hosts generally overlaps with the locus of continuous star formation
histories, in broad agreement with the situation for most inactive galaxies
(cf. Figure~3 of Kauffmann et~al. 2003a). This finding is consistent with the
idea that the evolution of the host galaxies of low-mass BHs is governed
principally by secular evolution (Greene et al. 2008; Jiang et~al. 2011).

\section{Summary}
Large-scale optical spectroscopic surveys remain the most effective
way to carry out a direct census of active massive BHs over a broad
range of masses and accretion rates.  We have developed an effective
automated selection procedure to select broad-line AGNs from the
SDSS spectral archive, with special emphasis on building a uniform
sample of AGNs with low BH masses that is more complete toward the
faint end of the luminosity function.  Particular care is given to
AGN-galaxy spectral decomposition and emission-line deblending.
Applying our methodology to the SDSS DR4, we compile a new sample of
309 broad-line AGNs with $\mbh < 2 \times 10^6$~\msun, drawn from a
parent sample of 8862 sources at $z<0.35$. The BH masses are
estimated from the luminosity and the width of the broad H$\alpha$
line, using the virial mass formalism of Greene \& Ho (2005b,
2007b). In detail, significant differences exist between the sample
of low-mass AGNs compiled here and that published by Greene \& Ho
(2007b); many are discovered for the first time. The BH masses span
the range $8 \times 10^4 - 2 \times 10^6$~\msun, with a median of
$1.2 \times 10^6$~\msun. The broad \ha\ luminosities range from
$10^{38.5}$  to $10^{42.3}$~\lum, and FWHMs from 490 to 2380~\kms\
(corrected for instrumental broadening). The Eddington ratios  range
from $\lesssim0.01$ to $\sim 1$, and 89 (30\%) objects have $\lratio
\leqslant 0.1$. Compared to the previous, most comprehensive catalog
of low-mass AGNs (Greene \& Ho 2007b), our sample contains more
systems accreting at a low Eddington rate ($\lratio < 0.1$).

We present some initial statistical results of this new
low-mass AGN sample, focusing on basic properties of the emission lines and
host galaxies that can be easily ascertained from the SDSS database. The host
galaxies have $g$-band luminosities from
$-22.2$ to $-15.9$~mag, with a median value comparable to $L^*$, and have
$u-g$ and $g-r$ colors typical of Sbc galaxies. %
%%... \wm{one sentence on morphology based on colors..?}%
We analyze $D_{4000}$ and H$\delta_{\rm A}$, two stellar indices sensitive to
age, and conclude that the majority of the host galaxies (174 out of 262) have
mean stellar ages less than 1~Gyr.  In general the hosts of low-mass AGNs
have experienced a history of continuous star formation, consistent
with the proposition, based on detailed analysis of their morphologies
(Jiang et al. 2011), that these galaxies evolve mainly through
secular processes.

Considering the selection effects that seriously act against finding
low-mass BHs at low accreting rates, our result implies that a
significant number of such low-mass BHs likely exist at the centers
of galaxies in the local Universe. This is also supported by the
detections of low-luminosity AGNs in very late-type disk galaxies in
the X-ray or infrared bands (see, e.g., Desroches \& Ho 2009). One
interesting question is whether there exists a lower limit to the
mass of massive black holes at the centers of galaxies. Our result
indicates that, if indeed there is any, this limit should be below a
few times $10^5$~\msun\ at least. Future deeper optical
spectroscopic surveys, such as the bigBOSS project, may shed new
light on the answer of this question by pushing the BH mass limit
even lower.

%%%%%%%%%%%%%%%%%%%%%%%%%%%%%%%%%%%%%%%%%%%%%%%%%%%%%%%%%%%%
\acknowledgments %
We thank the anonymous referee for his/her careful and
helpful comments. %
We thank Jian-Guo Wang and Kai Zhang for help in
improving the IDL figures, and Linhua Jiang, Peng Jiang and Hong-Lin
Lu for help with the MMT observations and data reductions.
\mbox{X.-B.\,D.} thanks Huiyuan Wang and Hong-Guang Shan for
valuable discussions, Yifei Chen and Paul Collison for computing
support, and Luis Ho for supporting his visit to Carnegie
Observatories (after March 2011). This work is supported by
Chinese NSF grants NSF-10703006, %xbdong-imbh
NSF-10973013, %tinggui-partial_2
NSF-11033007, %weimin-key_2
and NSF-11073019, %xbdong-qso
the SOAC project CHINARE2012-02-03, %polar
a National 973 Project of China (2009CB824800), %JF Lu--xbdong
and the Fundamental Research Funds for the Central Universities
(USTC WK2030220004). The research of \mbox{L.C.H.} is supported by
the Carnegie Institution for Science. %
XF acknowledges support from NSF Grant AST 08-06861, and from a
Packard Fellowship for Science and Engineering. %
Funding for the SDSS and SDSS-II has been provided by the Alfred P.
Sloan Foundation, the Participating Institutions, the National
Science Foundation, the U.S. Department of Energy, the National
Aeronautics and Space Administration, the Japanese Monbukagakusho,
the Max Planck Society, and the Higher Education Funding Council for
England.  The SDSS Web Site is http://www.sdss.org/. The SDSS is
managed by the Astrophysical Research Consortium for the
Participating Institutions. The Participating Institutions are the
American Museum of Natural History, Astrophysical Institute Potsdam,
University of Basel, University of Cambridge, Case Western Reserve
University, University of Chicago, Drexel University, Fermilab, the
Institute for Advanced Study, the Japan Participation Group, Johns
Hopkins University, the Joint Institute for Nuclear Astrophysics,
the Kavli Institute for Particle Astrophysics and Cosmology, the
Korean Scientist Group, the Chinese Academy of Sciences (LAMOST),
Los Alamos National Laboratory, the Max-Planck-Institute for
Astronomy (MPIA), the Max-Planck-Institute for Astrophysics (MPA),
New Mexico State University, Ohio State University, University of
Pittsburgh, University of Portsmouth, Princeton University, the
United States Naval Observatory, and the University of Washington.
%%%%%%%%%%%%%%%%%%%%%%%%%%%%%%%%%%%%%%%%%%%%%%%%%%%%%%%%%%%%

%%%%%%%%%%%%%%%%%%%%%%%%%%%%%%%%%%%%%%%%%%%%%%%%

\clearpage
%%%%%%%%%%%%%%%%%%%%%%%%%%%%%%%%%%%%%%%%%%%%%%%%%%%%%%%%

\setcounter{figure}{0}
\setcounter{table}{0}

%%%%%%%%%%%%%%%%%%%%%%%%%% figures  %%%%%%%%%%%%%%%%%%%%%%

\begin{figure}[tbp]
\epsscale{1} \plotone{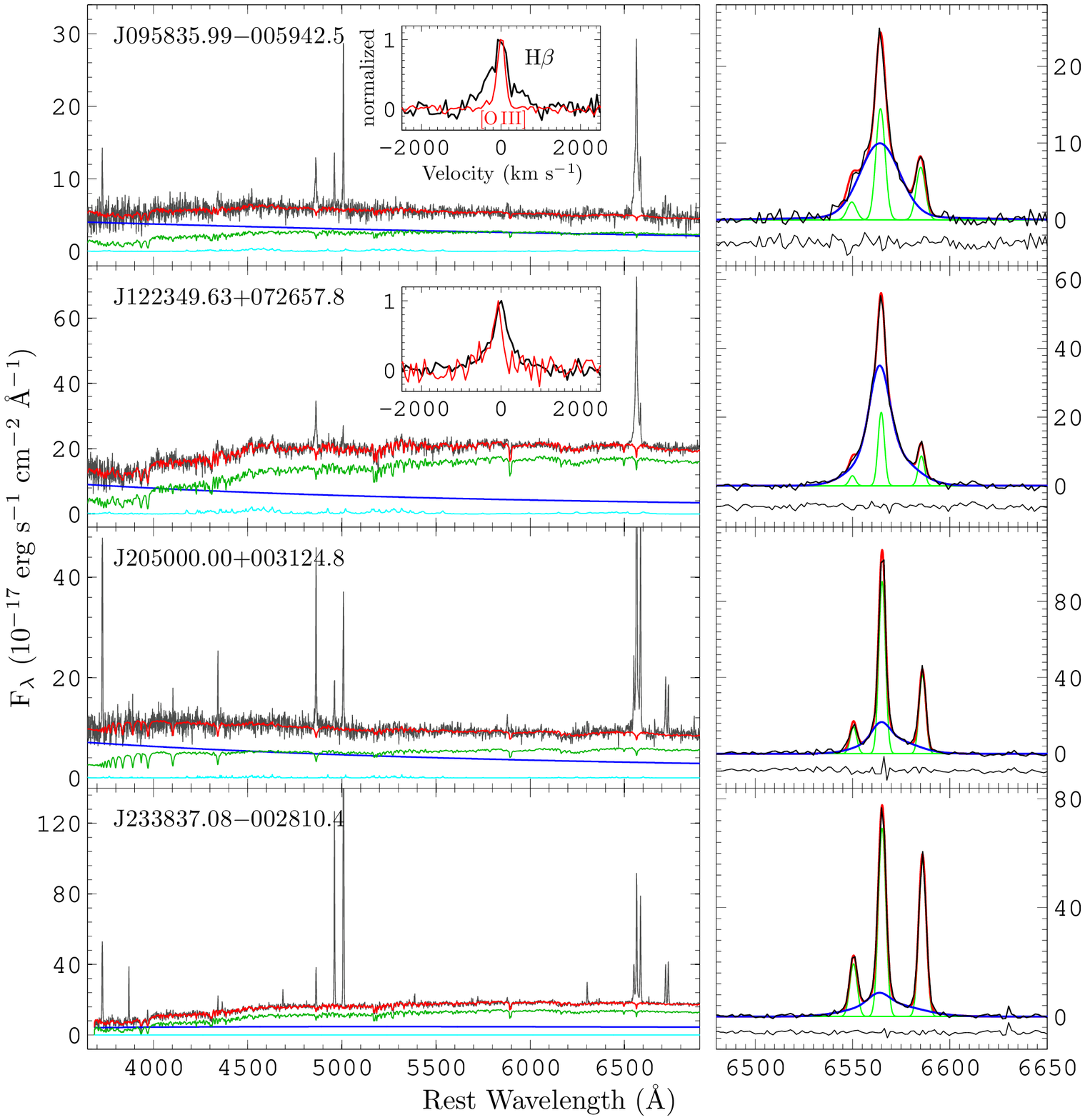}
\caption{\label{fig-demo4}%
%%Demonstration
Illustration of the continuum and emission-line fittings for four
low-accretion, low-mass Seyfert 1s that are not included in the
Greene \& Ho (2007b) sample. %
{\it Left}: Observed SDSS spectrum (black), the total model
(red), the decomposed components of the host galaxy (green), the AGN
continuum (blue), and the \feii\ multiplet emission (cyan). %
The inset is a zoomed-in view of a comparison of the
\hb\ line (black) and the \oiii\,$\lambda5007$ line profile (ref)
in the velocity space ($x$-axis), both with the continuum subtracted and the
peak flux density normalized at unity.
{\it Right}: Emission-line profile fitting in the \ha~$+$~\nii\ region.
}
\end{figure}
%%%%%%%%%%%%%%%%%%%%%%%%%%%%%%%%

\begin{figure}[tbp]
\epsscale{1} \plotone{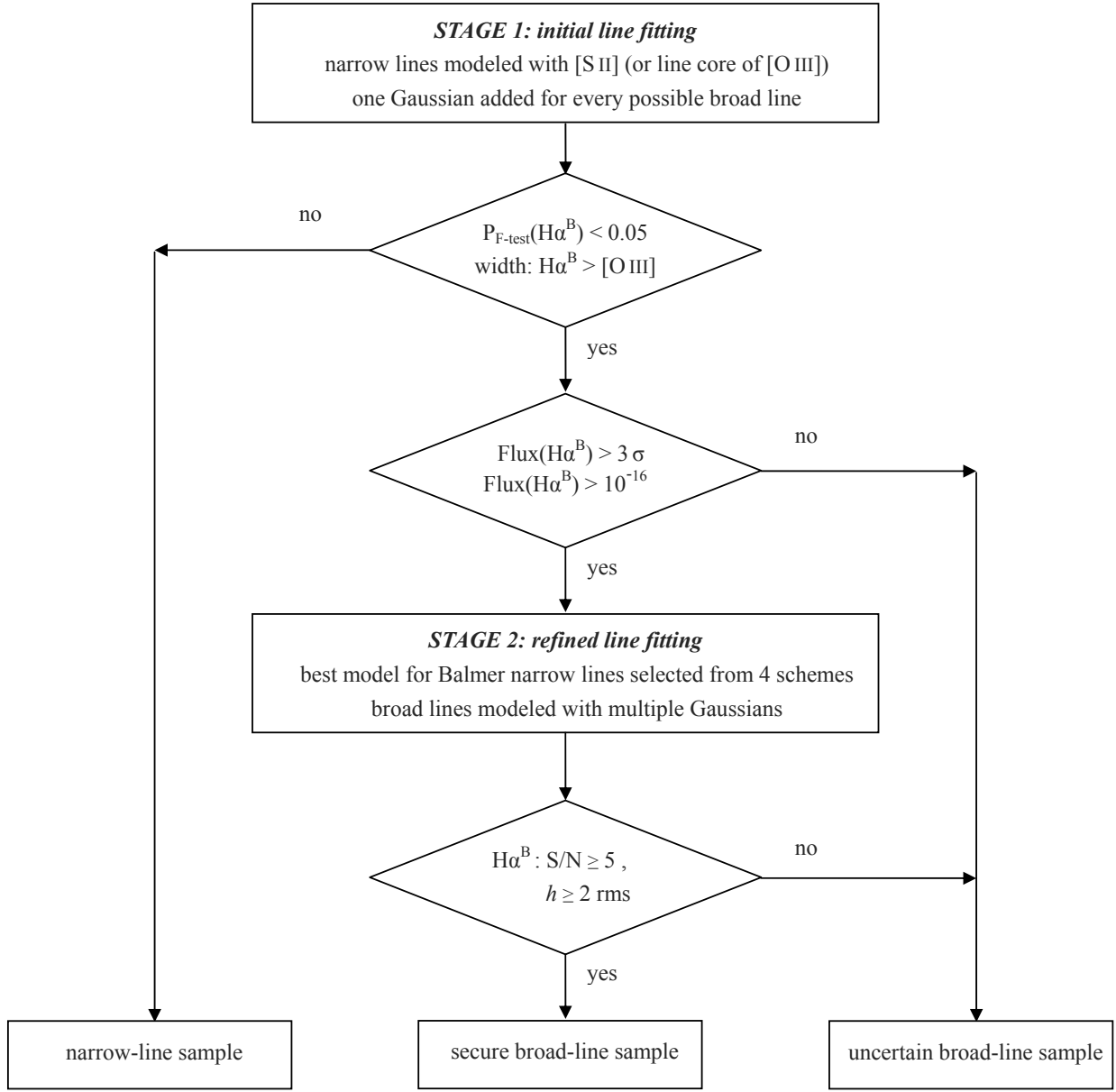}
\caption{\label{fig-flowchart}%
Flow chat of the automated procedure to select objects with the broad
\ha\ line. See Section~3 for details.}
\end{figure}

\begin{figure}[tbp]
\epsscale{1.1} \plottwo{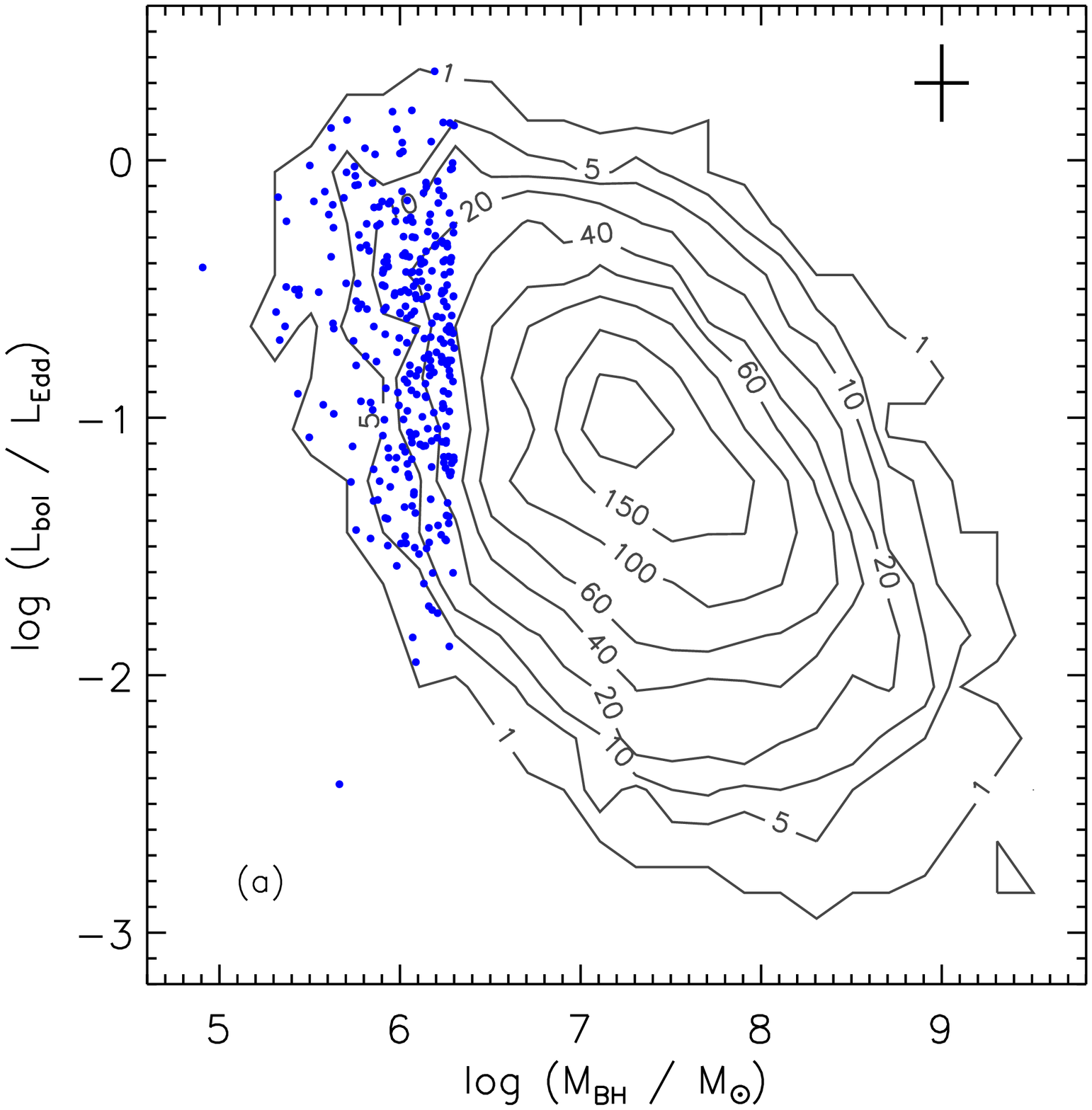}{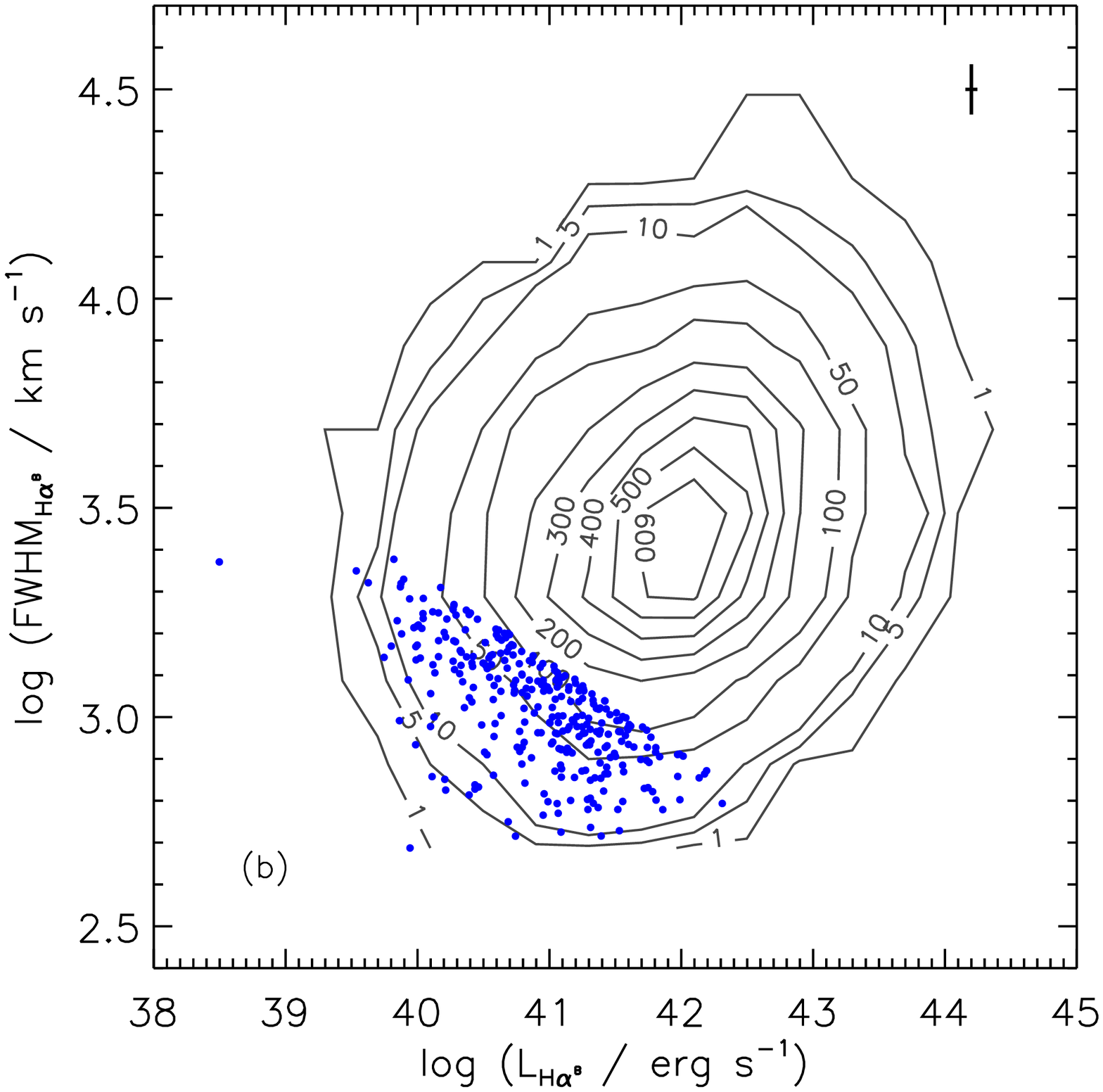}
\caption{\label{fig-our4dist}%
Distribution of our
low-mass BH sample (blue dots)
in the \lratio\ vs. \mbh\ plane (panel a) and the
FWHM vs. luminosity plane for the broad \ha\ line (panel b),
respectively. Over-plotted are
the contours of the parent 8862 broad-line AGNs
at $z < 0.35$ uniformly selected from the SDSS DR4.
The top-right corner of each panel shows a representative error bar,
the length of which corresponds to the 1 $\sigma$ total error
(see Section 2.3).}
\end{figure}

\begin{figure}[tbp]
\epsscale{0.8} \plotone{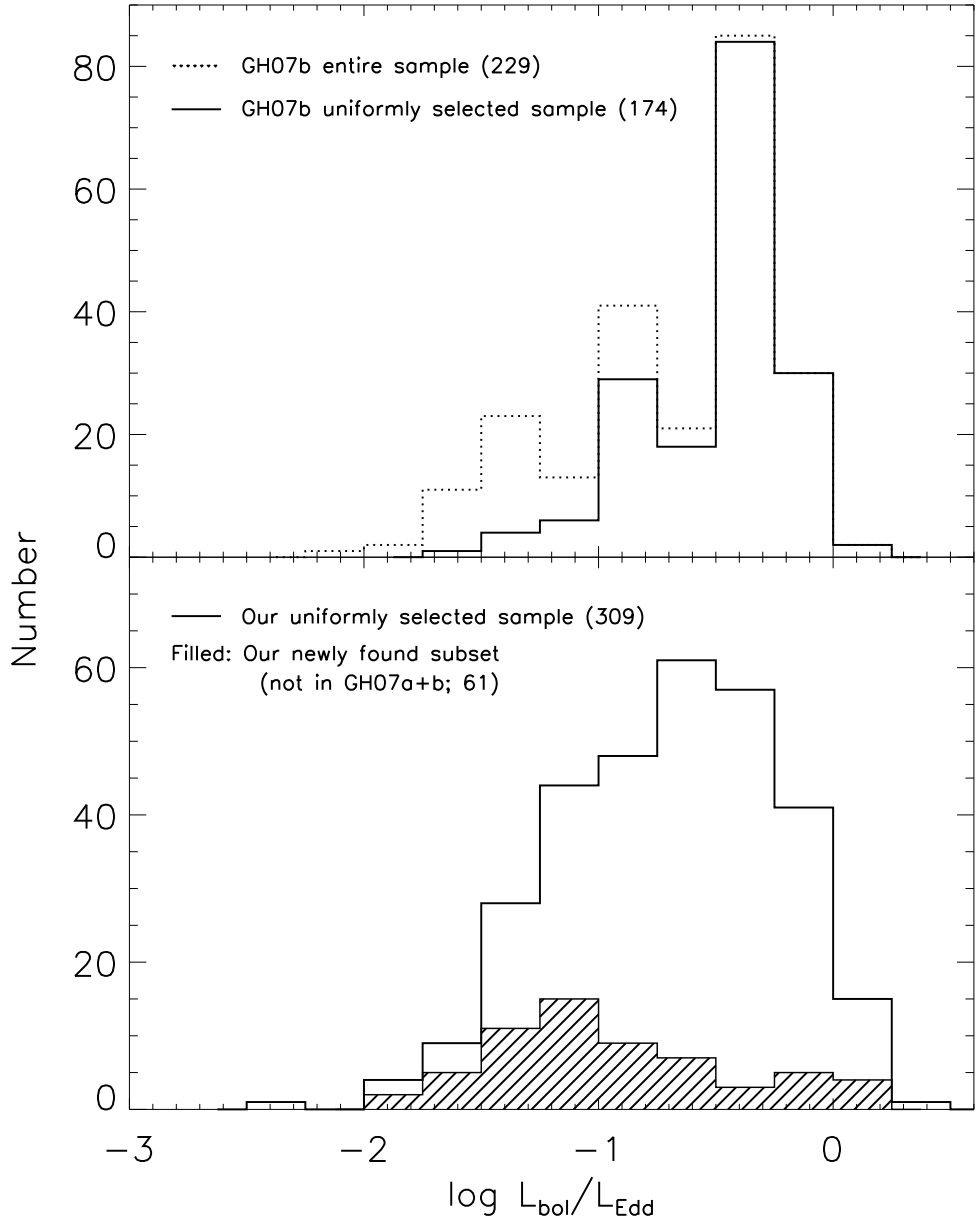}
\caption{\label{fig-compgh07ell}%
Comparison of the \lratio\ distribution of our low-mass BH sample
with that of Greene \& Ho (2007b).}
\end{figure}

\begin{figure}[tbp]
\epsscale{1} \plotone{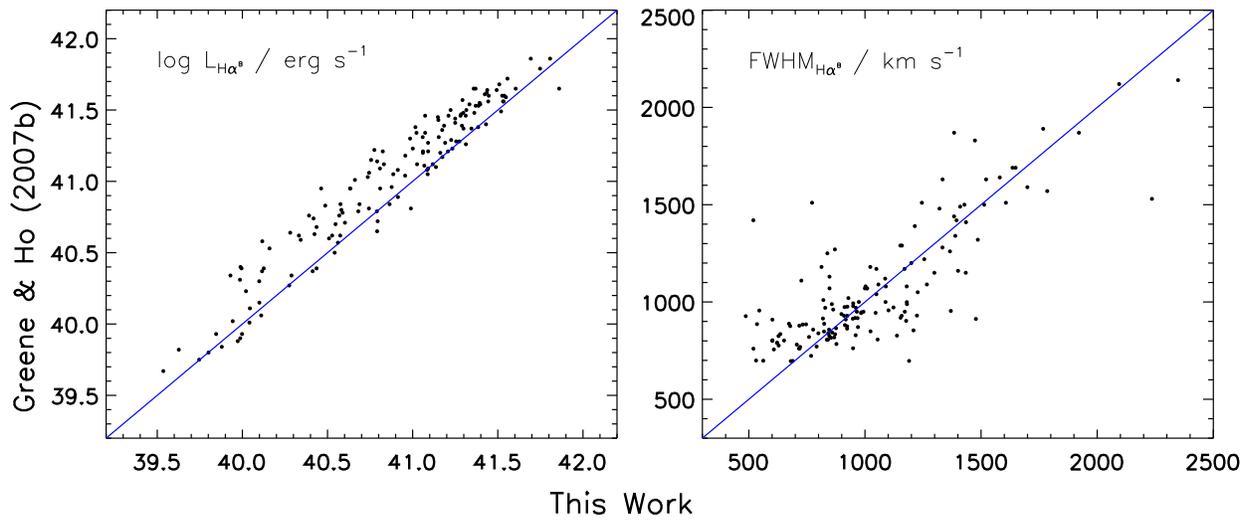}
\caption{\label{fig-common149}%
Comparison of the luminosities and FWHMs of broad \ha\ for the
149  objects in common between our sample and that of Greene \& Ho (2007b).}
\end{figure}

\begin{figure}[tbp]
\epsscale{1} \plotone{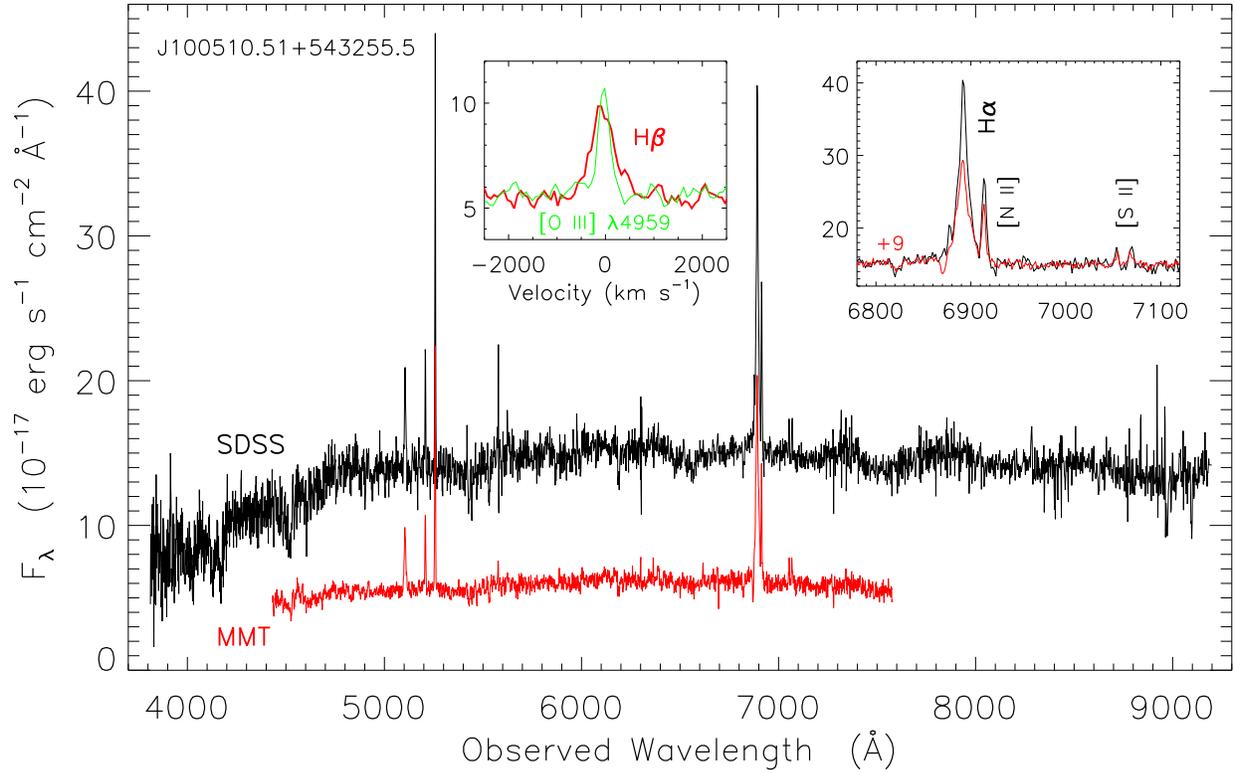}
\caption{\label{fig-mmt}%
Demonstration of the spectra of a newly discovered Seyfert 1 galaxy with a
low-mass BH and a low accretion rate, taken
by the SDSS (black) and MMT (red), respectively.
The inset on the left shows a zoomed-in view of the \hb\ line
region in the MMT spectrum
(without continuum subtraction), in direct comparison with the
\oiii\,$\lambda4959$ line profile in the velocity space ($x$-axis).
The inset on the right shows a zoomed-in view of the
\ha\,$+$\,\nii\,$+$\,\sii\ region of the SDSS spectrum (black) and the
MMT spectrum (red).}
\end{figure}

\clearpage

\begin{figure}[tbp]
\epsscale{1} \plotone{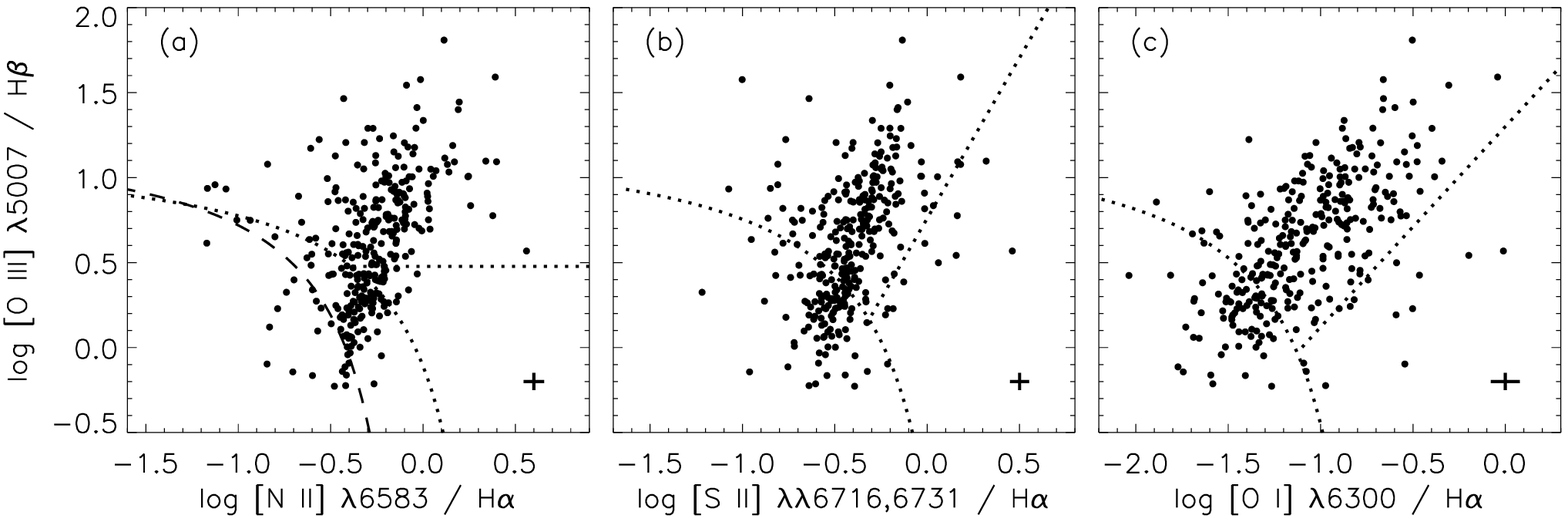} %
\caption{\label{fig-bpt}%
Narrow-line diagnostic diagrams of \oiii\,$\lambda 5007$/\hb\ vs.
\nii\,$\lambda 6583$/\ha\ (a), vs. \sii\,$\lambda\lambda
6716,6731$/\ha\ (b), and vs. \oi\,$\lambda 6300$/\ha\ (c) for
our low-mass BH AGN sample. The dotted, curved
lines separating \hii\ regions, AGNs and LINERs
are taken from Kewley et~al. (2001) and Kewley et~al. (2006), respectively.
In panel (a), the dashed line is an
empirical demarcation between \hii\ regions and AGNs given by
Kauffmann et~al. (2003), and the dotted horizontal line corresponds
to \oiii\,$\lambda 5007$/\hb\ $=3$, a traditional line separating
AGN and LINERs. The symbols are the same as in
Figure~\ref{fig-our4dist}.}
\end{figure}

\begin{figure}[tbp]
\epsscale{1.1} \plottwo{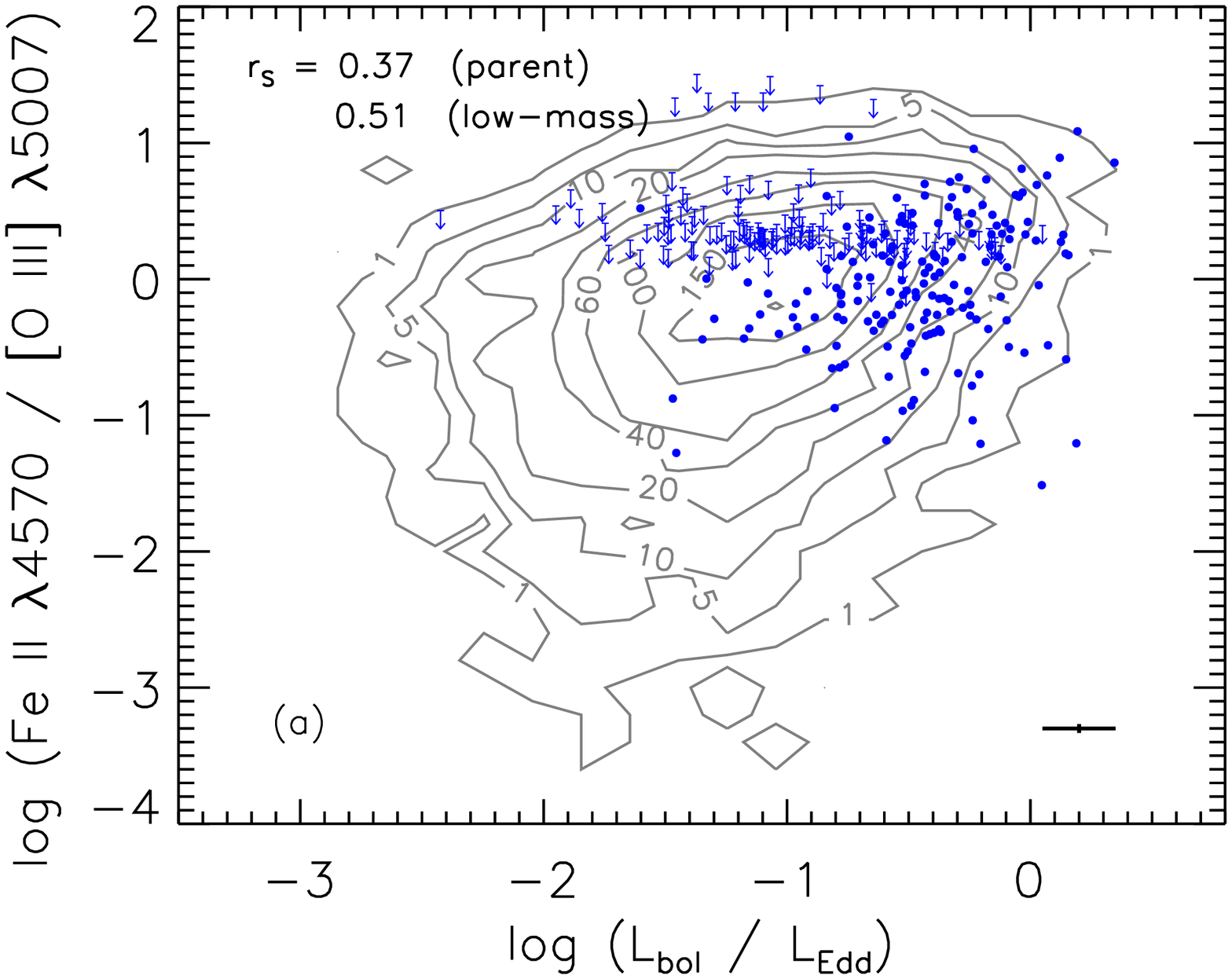}{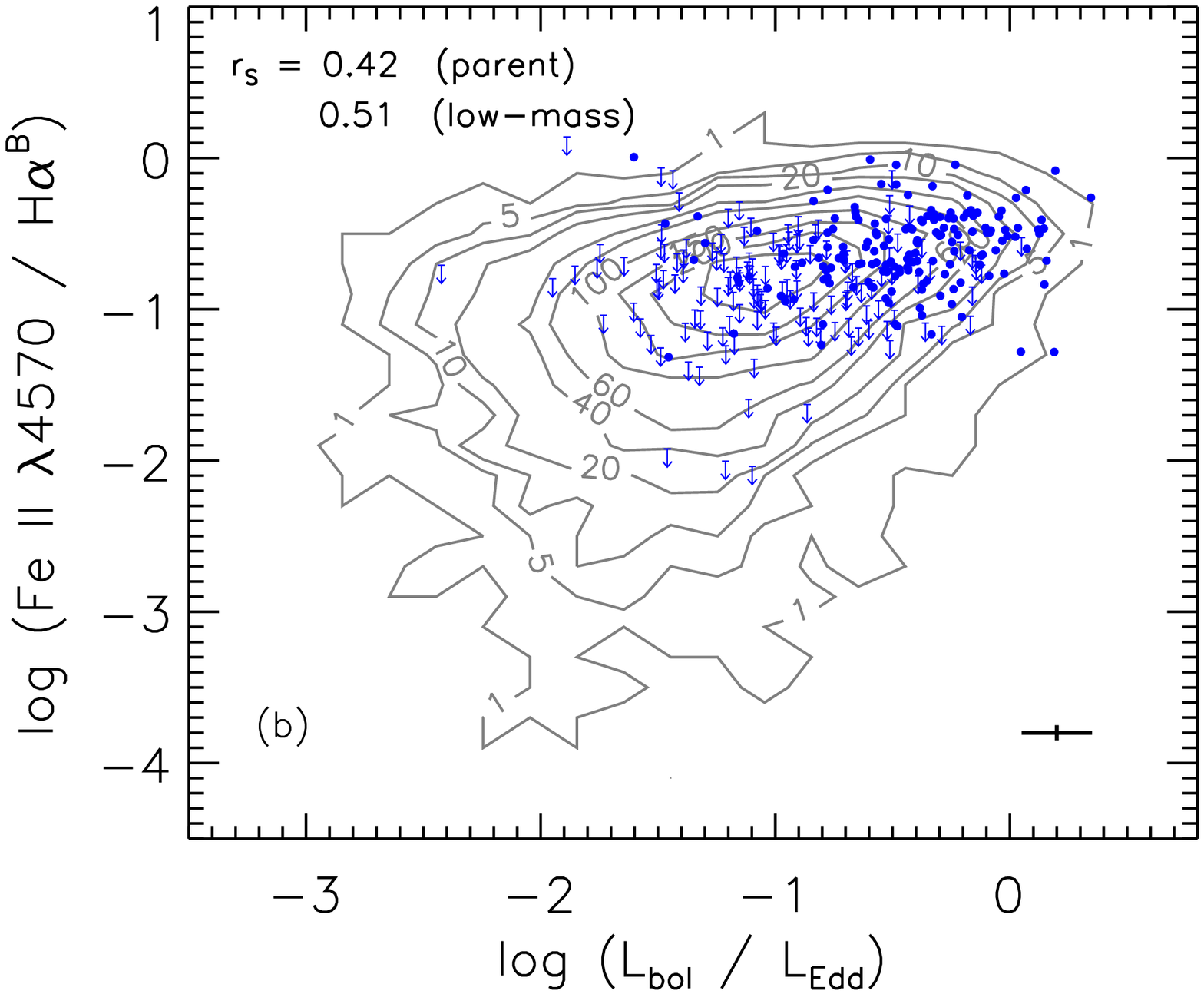}
\caption{\label{fig-pc1}%
Correlations of the Eddington ratio with the line ratios,
\feii\,$\lambda 4570$/\oiii\,$\lambda 5007$ (panel a) and
\feii\,$\lambda 4570$/(broad \ha) (panel b), respectively.
The contours
are for the entire parent broad-line AGN sample and the blue dots
are for the present low-mass AGN sample.
Also denoted are the Spearman
correlation coefficient \rs\ for the two samples.
The symbols are the same as in
Figure~\ref{fig-our4dist}.}
\end{figure}

\clearpage

\begin{figure}[tbp]
\epsscale{0.5} \plotone{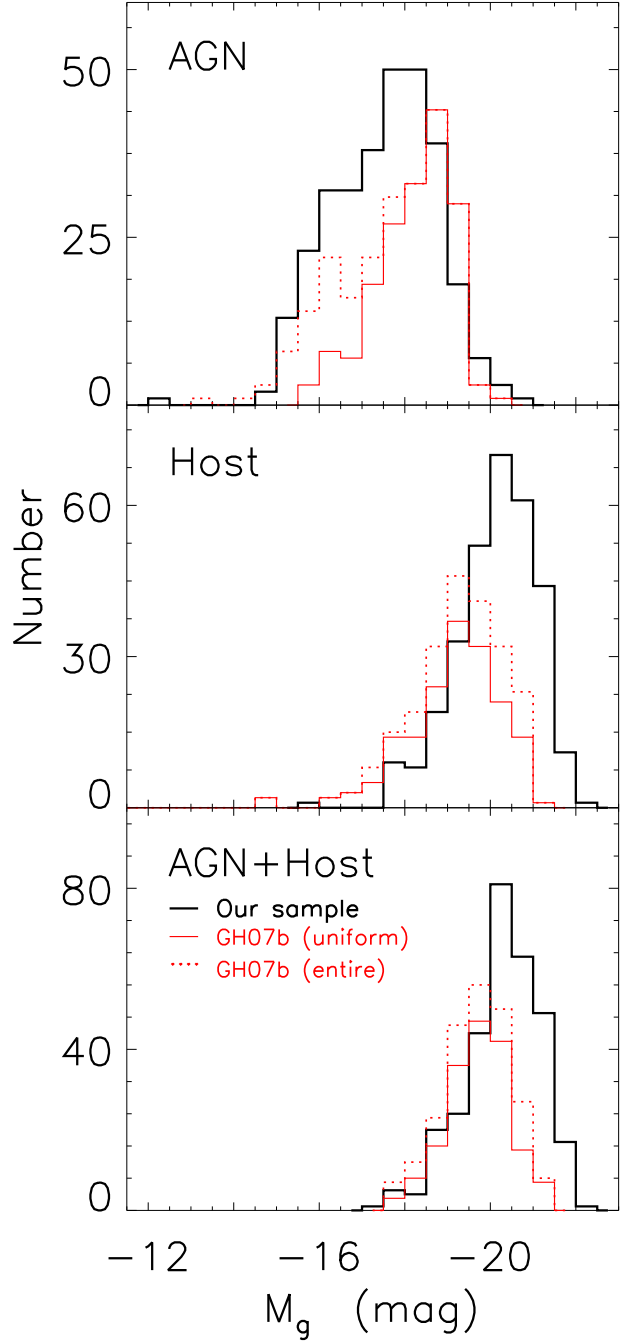} %
\caption{\label{fig-mags}%
Distributions of the absolute magnitudes in the $g$-band
pertaining to the AGN only ({\it top}), the host galaxy ({\it
middle}), and the total (AGN and host galaxy) ({\it bottom}),
for this sample (black solid line),
the uniformly selected
sample of Greene \& Ho (2007) (green solid lines)
and their entire sample (green dotted line).
The AGN contribution
is estimated from the broad \ha\ luminosity using the $L_{\rm
H\alpha}$--$L_{5100}$ relation of Greene \& Ho (2005[b]) and assuming a
power-law continuum slope $\beta = -1.56$ ($f_{\lambda} \propto
\lambda^\beta$; Vanden~Berk et al. 2001). }
\end{figure}

%
%\begin{figure}[tbp]
%\epsscale{1}
%\plottwo{J0838color.eps}{J0838sbp.eps}%%{J0838sbp_linear.eps} %
%\caption{\label{fig-J0838}%
%{\it Left}: The $gri$ three-color composite image of SDSSJ083803.67+540642.2
%($z=0.0295$). The red line marks a scale of 10\arcsec.
%A faint blue outer ring can be seen at about 10\arcsec\ away
%from the center. %
%{\it Right}: The $r$-band surface brightness profile (plus $\pm
%1$ $\sigma$ error bar) with sky subtracted and the fitting result.
%The best fit (red) is composed of PSF (blue) + \sersic\ (purple) +
%exponential disk (green). The PSF is extracted from a nearby star.
%In the fitting, the ring region (7--13\arcsec) has been masked out,
%and the galactic components (\sersic\ and exponential) have been
%convolved with the PSF. The cyan dotted line denotes the limiting
%surface brightness of 27.06~$\rm mag~arcsec^{-2}$. }
%\end{figure}

\begin{figure}[tbp]
\epsscale{1} \plotone{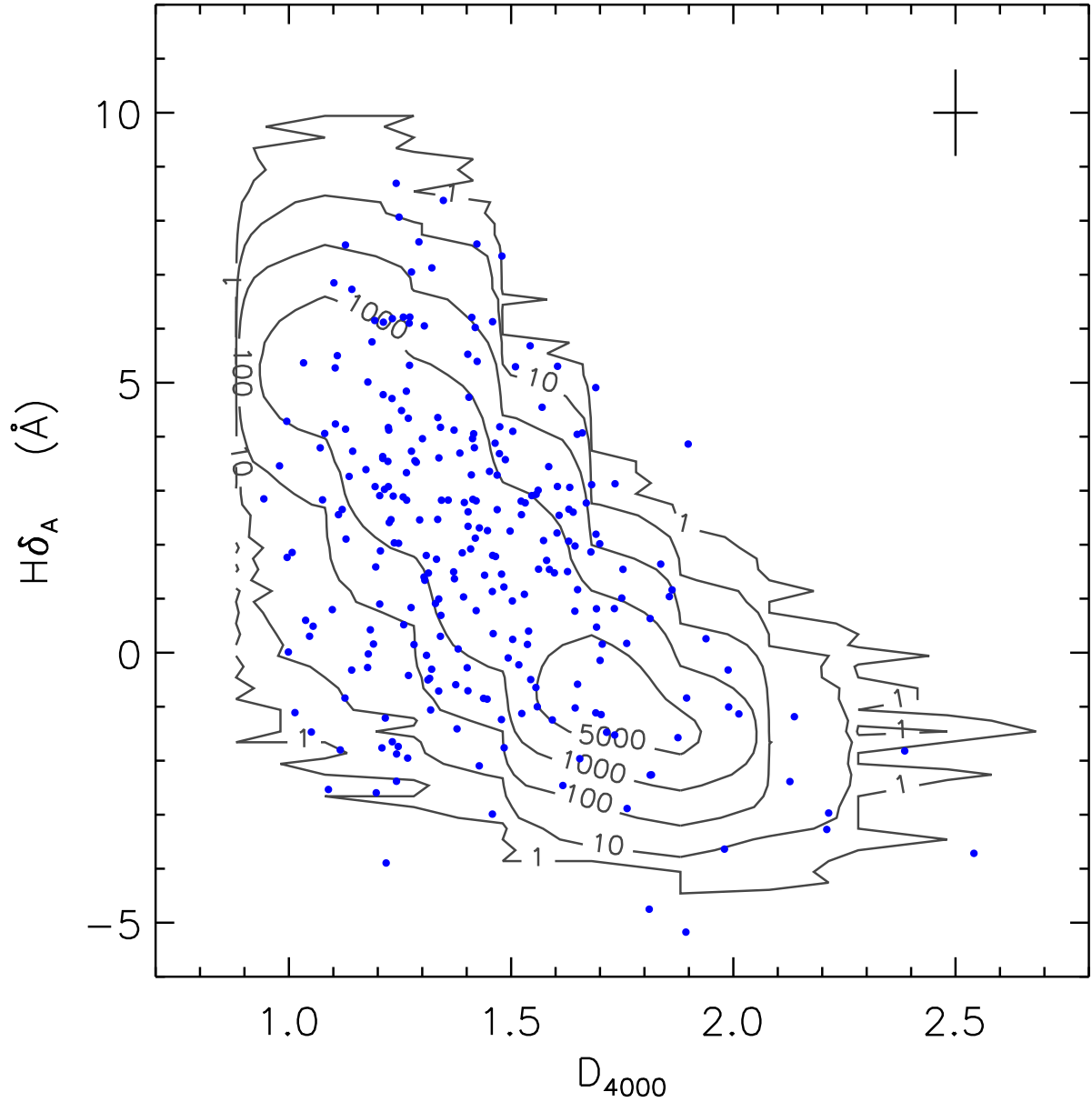} %
\caption{\label{fig-d4k}%
Distribution of the 262 low-mass BH host galaxies (blue dots) in
the plane of two stellar indexes, the 4000~\AA\ break and the
equivalent width of the H$\delta$ absorption. These objects have an AGN
contribution less than 75\% at 4000~\AA\ in their SDSS spectra. For
comparison the distribution of 318,500 inactive galaxies in the SDSS
DR4 is also plotted in contours. The symbols are the same as in
Figure~\ref{fig-our4dist}.}
\end{figure}

\clearpage
%%%%%%%%%%%%%%%%%%%%%%%%%%%%%%%%%%%%%%%%%%%%%%%%%%%%%%%%%%%%%%%%%%%%%%%%
\begin{deluxetable}{rclccc}
\tablecolumns{6} \tabletypesize{\scriptsize} \tablewidth{0pc}
\label{tab-basic}
\tablecaption{The SDSS Sample} %
\tablehead{ \colhead{ID}  &
\colhead{SDSS Name} & \colhead{$z$} &
\colhead{$g$} & \colhead{$g-r$}     & \colhead{$A_{g}$} \\
\colhead{(1)} & \colhead{(2)} & \colhead{(3)} &
\colhead{(4)} & \colhead{(5)} & \colhead{(6)} }
\startdata
%ID & NAME                  & z      & PTROg & g-r  & A_g        % SPEC
1   & J000111.15$-$100155.7 & 0.0489 & 18.65 & 0.78 & 0.15   \\  % 52143-0650-174
2   & J000308.48$+$154842.2 & 0.1175 & 18.31 & 0.55 & 0.14   \\  % 52235-0750-609
3   & J001728.84$-$001826.8 & 0.1118 & 17.85 & 0.74 & 0.10       % 51795-0389-120
\enddata
\tablecomments{
Col. (1): Identification number assigned in this paper.
Col. (2): Official SDSS name in J2000.0.
Col. (3): Redshift measured by the SDSS pipeline.
Col. (4): Petrosian $g$ magnitude, uncorrected for Galactic extinction.
Col. (5): Petrosian $g-r$ color.
Col. (6): Galactic extinction in the $g$ band.
{\it (This table is available in its entirety in a machine-readable form in the online
journal. A portion is shown here for guidance regarding its form and content.)}
}
\end{deluxetable}

%%%%%%%%%%%%%%%%%%%%%%%%%%%%%%%%%%%%%%%%%%%%%%%%%%%%%%%%%%%%%%%%%%%%%

%%%%%%%%%%%%%%%%%%%%%%%%%%%%%%%%%%%%%%%%%%%%%%%%%%%%%%%%%%%%%%%%%%%
%\begin{deluxetable}{rcccccccccccccc}
%\tablenum{2}
%\tablecolumns{15} \tablewidth{0pc}
%\tabletypesize{\tiny}

\begin{sidewaystable*}
\topmargin 0.0cm \evensidemargin = 0mm \oddsidemargin = 0mm
\scriptsize
\label{tab-emline}
%\tablecaption{Emission-line Measurements}
\caption{Emission-line Measurements}
\begin{tabular}{rcccccccccccccc}
 \hline \hline
%\tablehead{
%\colhead{ID} &
%\colhead{[O\,{\tiny II}] $\lambda3727$} &
%\colhead{Fe\,{\tiny II} $\lambda4570$} &
%\colhead{H$\beta^{\rm N}$} &
%\colhead{H$\beta^{\rm B}$} &
%\colhead{[O\,{\tiny III}] $\lambda5007$} &
%\colhead{[O\,{\tiny I}] $\lambda6300$} &
%\colhead{H$\alpha^{\rm N}$} &
%\colhead{H$\alpha^{\rm B}$} &
%\colhead{[N\,{\tiny II}] $\lambda6583$} &
%\colhead{[S\,{\tiny II}] $\lambda6716$} &
%\colhead{[S\,{\tiny II}] $\lambda6731$} &
%\colhead{FWHM$_{\rm H\alpha^B}$}  &
%\colhead{FWHM$_{\rm [O\,III]}$}   &
%\colhead{FWHM$_{\rm [S\,II]}$}    \\
%\colhead{(1)} &
%\colhead{(2)} &
%\colhead{(3)} &
%\colhead{(4)} &
%\colhead{(5)} &
%\colhead{(6)} &
%\colhead{(7)} &
%\colhead{(8)} &
%\colhead{(9)} &
%\colhead{(10)} &
%\colhead{(11)} &
%\colhead{(12)} &
%\colhead{(13)} &
%\colhead{(14)} &
%\colhead{(15)}  }
ID &
[O\,{\tiny II}] $\lambda3727$ &
Fe\,{\tiny II} $\lambda4570$  &
H$\beta^{\rm N}$ &
H$\beta^{\rm B}$ &
[O\,{\tiny III}] $\lambda5007$ &
[O\,{\tiny I}] $\lambda6300$   &
H$\alpha^{\rm N}$ &
H$\alpha^{\rm B}$ &
[N\,{\tiny II}] $\lambda6583$ &
[S\,{\tiny II}] $\lambda6716$ &
[S\,{\tiny II}] $\lambda6731$ &
FWHM$_{\rm H\alpha^B}$  &
FWHM$_{\rm [O\,III]}$   &
FWHM$_{\rm [S\,II]}$    \\
(1) & (2)  & (3)  & (4)  & (5)  & (6)  & (7) & (8) &
(9) & (10) & (11) & (12) & (13) & (14) & (15)  \\
\hline
%\startdata
%%ID&   LOGF_3727  &   LOGF_4570  &   LOGF_NHB  &   LOGF_BHB  &    LOGF_5007  &   LOGF_6300  &   LOGF_NHA  &    LOGF_BHA  &   LOGF_6583  &   LOGF_6716  &   LOGF_6731  &  FWHM_BHA  &  FWHM_OIII  &  FWHM_SII      %  SPEC
1   &    $-$15.12  &   $<-$15.45  &   $-$15.39  &  $<-$15.76  &     $-$14.79  &    $-$15.84  &   $-$14.65  &    $-$14.71  &    $-$15.08  &    $-$15.28  &    $-$15.46  &      1921  &        215  &       195  \\  %  52143-0650-174
2   &    $-$15.32  &    $-$15.19  &   $-$15.61  &   $-$15.34  &     $-$15.44  &    $-$16.11  &   $-$15.08  &    $-$14.78  &    $-$15.51  &    $-$15.74  &    $-$15.87  &       924  &        164  &       158  \\  %  52235-0750-609
3   &    $-$15.34  &   $<-$15.38  &   $-$15.40  &  $<-$15.81  &     $-$15.13  &    $-$16.12  &   $-$14.74  &    $-$14.62  &    $-$14.95  &    $-$15.49  &    $-$15.53  &      1022  &        168  &       256  \\  %  51795-0389-120
%\enddata
%\tablecomments{\normalsize
\hline
\end{tabular}
\medskip
\vfill
{\normalsize Note. ---
Col. (1): Identification number assigned in this paper. %
Cols. (2)--(12): Emission-line fluxes (or 3 $\sigma$ upper limits) in
log-scale, in units of $\mathrm{erg~s^{-1}~cm^{-2}}$. Note that
these are observed values; no NLR or BLR extinction correction has
been applied. The superscripts ``N'' and ``B'' in cols. 4, 5, 8, 9
and 13 refer to the narrow and broad components of the line, respectively. %
Cols. (13)--(15): Line widths (FWHM) that are calculated from the
best-fit models and have been corrected for instrumental broadening
using the values measured from arc spectra and tabulated by the
SDSS; in units of \kms. %
{\it (This table is available in its entirety in a machine-readable
form in the online journal. A portion is shown here for guidance
regarding its form and content.)} }
%\end{deluxetable}
\end{sidewaystable*}
\clearpage
%%%%%%%%%%%%%%%%%%%%%%%%%%%%%%%%%%%%%%%%%%%%%%%%%%%%%%%%%%%%%%%%%%%%%%%%

\begin{deluxetable}{rcccccc}
%\tablenum{3}
\tablecolumns{7} \tabletypesize{\scriptsize}
\tablewidth{0pc} %
\tablecaption{\label{tab-lumin}Luminosity and Mass Measurements} %
\tablehead{
\colhead{ID} &
\colhead{$M_g$(total)} & \colhead{$M_g$(AGN)} & \colhead{$M_g$(host)} & %
\colhead{log $L_{\rm H\alpha^B}$} & \colhead{log $M_{\rm BH}$} & %
\colhead{log $L_{\rm bol}/L_{\rm Edd}$} \\
\colhead{(1)} &
\colhead{(2)} &
\colhead{(3)} &
\colhead{(4)} &
\colhead{(5)} &
\colhead{(6)} &
\colhead{(7)} }
\startdata
%% ID&   TOTAL    &     AGN    &    HOST    &   LBHA  &  MBH  &   ELL        %  SPEC
1    &  $-$18.30  &  $-$15.65  &  $-$18.20  &  40.04  &  6.2  &  $-$1.6  \\  %  52143-0650-174
2    &  $-$20.66  &  $-$17.23  &  $-$20.61  &  40.77  &  5.9  &  $-$0.6  \\  %  52235-0750-609
3    &  $-$21.06  &  $-$17.47  &  $-$21.02  &  40.89  &  6.0  &  $-$0.7      %  51795-0389-120
\enddata
\tablecomments{ Col. (1): Identification number assigned in this paper. %
Col. (2): Total $g$-band absolute magnitude. %
Col. (3): AGN $g$-band absolute magnitude, estimated from $L_{\rm
H\alpha^B}$ given in col. (5) and a conversion from $L_{\rm
H\alpha}$ to $M_g$ assuming $f_{\lambda} \propto \lambda^{-1.56}$. %
Col. (4): Host galaxy $g$-band absolute magnitude, obtained by
subtracting the AGN luminosity from the total luminosity. %
Col. (5): Luminosity of broad H$\alpha$, in units of \lum. %
Col. (6): Virial mass estimate of the BH, in units of $M_{\odot}$. %
Col. (7): Eddington ratio. {\it (This table is available in its
entirety in a machine-readable form in the online journal. A portion
is shown here for guidance regarding its form and content.)} }
\end{deluxetable}

%%%%%%%%%%%%%%%%%%%%%%%%%%%%%%%%%%%%%%%%%%%%%
\begin{deluxetable}{lcccccccc}
\tablenum{4} \tabletypesize{\scriptsize} \tablewidth{0pt}
\tablecaption{\label{tab-compgh07}%
Comparison with the low-mass BH sample of Greene \& Ho (2007b)
\tablenotemark{~a} } %
\tablehead{ Sample \tablenotemark{~b}  & $N$ & log\,$z$ &
log\,$L_{\rm H\alpha^B}$ & log\,FWHM$_{\rm H\alpha^B}$ & log\,$\mbh$
& log\,$\lratio$ } %
\startdata
GH07b entire sample             & 229 & $-1.06$; 0.25  & 41.1; 0.6  & 3.0; 0.1 & 6.1; 0.2 & $-0.5$; 0.5  \\
GH07b uniformly selected sample & 174 & $-1.01$; 0.23  & 41.3; 0.4  & 3.0; 0.1 & 6.1; 0.2 & $-$0.4; 0.3  \\
our uniformly selected sample   & 309 & $-1.03$; 0.25  & 41.0; 0.6  & 3.0; 0.1 & 6.1; 0.2 & $-$0.7; 0.5  \\
our newly found subset          & 61  & $-1.06$; 0.27  & 40.5; 0.6  & 3.1; 0.2 & 6.1; 0.2 & $-$1.1; 0.5  %
\enddata

\tablenotetext{\it a}{~For each entry, we list the median value and
standard deviation.} %
\tablenotetext{\it b}{~GH07b entire sample comprises 229 objects,
including 174 objects uniformly selected according to the detection
threshold of Greene \& Ho (2007b) and 55 less secure candidates
manually picked up; GH07b uniformly selected sample comprises only
the 174 objects; our uniformly selected sample comprises 309 objects
selected according to our criteria; our newly found subset comprises
61 objects not included in the broad-line AGN sample of Greene \& Ho
(2007a) or the entire low-mass sample of Greene \& Ho (2007b).}
\end{deluxetable}

\end{document}